\documentclass[manuscript, screen]{acmart}
\usepackage{booktabs} 
\usepackage[utf8]{inputenc}

\usepackage[ruled]{algorithm2e} 
\usepackage{tabularx}
\usepackage{array}
\usepackage{ragged2e}
\usepackage{adjustbox}
\newcolumntype{P}[1]{>{\RaggedRight\hspace{0pt}}p{#1}}
\usepackage{booktabs}
\usepackage{rotating}
\usepackage{pgfplots,tikz,refcount,nameref,totcount,etoolbox}
\usepackage{etaremune}
\usepackage{csquotes}
\usepackage{paralist}
\usepackage{enumitem}

\pgfplotsset{compat=1.14}

\acmJournal{CSUR}
\acmVolume{9}
\acmNumber{4}
\acmArticle{39}
\acmYear{2018}
\acmMonth{3}

\setcopyright{acmcopyright}

\acmDOI{0000001.0000001}

\begin{document}
\title{State of the Software Development Life-Cycle for the Internet-of-Things}

\author{João Pedro Dias}
\orcid{1234-5678-9012-3456}
\affiliation{%
  \institution{Faculty of Enginnering, University of Porto}
  \streetaddress{Rua Dr. Roberto Frias}
  \city{Porto}
  \postcode{4200-465}
  \country{Portugal}}
\email{jpmdias@fe.up.pt}

\author{Hugo Sereno Ferreira}
\affiliation{%
  \institution{Faculty of Enginnering, University of Porto}
  \streetaddress{Rua Dr. Roberto Frias}
  \city{Porto}
  \postcode{4200-465}
  \country{Portugal}}
\email{hugosf@fe.up.pt}

\renewcommand\shortauthors{Zhou, G. et al}

\begin{abstract}

Software has a longstanding association with a state of crisis considering its success rate. The explosion of Internet-connected devices – Internet-of-Things – adds to the complexity of software systems. The particular characteristics of these systems, such as being large-scale and its heterogeneity, pose increasingly new challenges.  In this paper, we first briefly introduce the IoT paradigm and the current state of art of software development. Then, we delve into the particularities of developing software for IoT systems and systems of systems, given an overview of what are the current methodologies and tools for design, develop and test such systems. The findings are discussed, revealing open issues and research directions, and reveal that the nowadays IoT software development practices are still lagging behind of what are the current best practices.

\end{abstract}

%
%

\ccsdesc[500]{Software and its engineering~Software creation and management}
\ccsdesc[300]{Software and its engineering~Ultra-large-scale systems}
\ccsdesc[100]{Software and its engineering~Development frameworks and environments}
\ccsdesc[100]{Software and its engineering~Software testing and debugging}
\ccsdesc[300]{Computer systems organization~Embedded and cyber-physical systems}

%
%

\keywords{Software Engineering, Internet-of-Things}

\maketitle

\section{Introduction}

Software has a long-standing association with states of crisis considering its success rate.
From a developer’s perspective, developing software systems is hard, and developers have a
high-proneness for introducing bugs during the development of such systems.

The explosion of Internet-connected devices with computing capabilities, also known as
the Internet-of-Things (IoT) – the recent peak of ubiquitous connectivity and computing – adds to
the complexity of software systems. One of the keys successes of IoT depends on the ability to
seamlessly interconnect the plethora of available devices and services.

However, this comes with an increase in complexity that might further impact an already
low success rate of software project development. The very nature of IoT systems is characterized
by (1) being ultra-large-scale, (2) having highly-dynamic topologies, (3) being highly
heterogeneous, and (4) their multi-domain nature. Together, they pose increasingly new challenges
on how to design, develop and maintain them.

On this paper, we delve into an analysis on the current state of art of software engineering practices for the IoT, focusing in three phases of the software development life-cycle, namely, the design, develop and test software systems.

\section{Background}

\subsection{Internet-of-Things}

Internet-of-Things (IoT) is a paradigm that consists of the uniquely identifiable objects (things) and their virtual representations within the Internet structure~\cite{Barricelli2015}. Broadly, it refers to the inter-connectivity between ordinary devices alongside with the device contextual awareness, sensing capability, and autonomy~\cite{Hossain2015}. Global Standards Initiative on IoT (IoT-GSI) defines IoT as the \textit{infrastructure of the information society}~\cite{Kafle2016}, foreseeing the advance towards new \textit{smart spaces}~\cite{Korzun2013} by the means of ubiquitous computing~\cite{Friedemann2011}.

The IoT paradigm opens the doors to new innovations that will build a novel type of interactions among \textit{things} and humans, enabling the realization of smart cities, infrastructures, and services for enhancing the quality of life and utilization of resources. Thus, IoT envisions a new world of connected devices and humans in which the quality of life is enhanced because management of the city and its infrastructure is less cumbersome, health services are conveniently accessible, and disaster recovery is more efficient~\cite{Buyya2016}. 

From a technical point-of-view, one can consider that a major role of the IoT consists of the delivery of highly complex knowledge-based and action-oriented applications in real-time. In order to be able to reach such an end, several considerations should be done when considering the full life-cycle of this system, from conceptualization to development, from test to deployment and maintenance. These include, but are not limited to: development of scalable architectures, moving from closed systems to open systems, dealing with privacy and ethical issues (due to the involved in data sensing), heterogeneity support, data storage, data processing, decision-making, designing interaction protocols, autonomous management, communication protocols, smart objects and service discovery, programming frameworks and languages, resource management, data and network management, real-time necessities, power and energy management, governance and interoperability~\cite{Buyya2016}.

An intertwined concept with the IoT one is the Web of Things (WoT). WoT is a term used to describe approaches, software architectural styles and programming patterns that allow real-world objects to be part of the World Wide Web. In simple terms, similarly to what the Web (Application Layer) is to the Internet (Network Layer), the Web of Things provides an Application Layer that simplifies the creation of IoT applications.

\subsubsection{A Brief History \& Vision}

Almost five decades after the birth of the Internet by ARPANET\footnote{Initiated in 1969 by the Advanced Research Projects Agency (DARPA) of the Department of Defence (DoD) ARPANET was the first wide area packet switching network.}~\cite{Perry1988}, the term \textit{Internet} refers to the vast category of applications and protocols built on top of sophisticated and interconnected computer networks, available 24/7 and serving above 3.5 billion users worldwide circa 2016~\cite{IntSta17,Buyya2016}.

As of today, we can consider that ubiquitous computation (\textit{ubicomp}, pervasive computing) and ubiquitous connectivity is neither a dream or a challenge anymore. Ubiquitous computation is visible anytime and everywhere, by using any device, in any location, and in any format. Plus, with ubiquitous connectivity, connectivity is available to everyone and everything, everywhere, every time. 

As consequence of the pervasive computing, the focus has shifted, from the goal of connecting people and make computation available for all, towards a seamless integration of people and devices to converge the physical realm with human-made virtual environments. This phenomenon has been known as the  Internet-of-Things (IoT) \textit{utopia}.

From a historical viewpoint, the term Internet-of-things was coined by Kevin Ashton, circa 1999, during a presentation about supply-chain management and the use of Radio-Frequency Identification (RFID) technology to enable computers to observe, identify and understand the world (without the limitations of human-entered data)\footnote{``I (Kevin Ashton) could be wrong, but I'm fairly sure the phrase \textit{Internet of Things} started life as the title of a presentation I made at Procter\& Gamble(P\&G) in 1999''~\cite{AshtonKevin2009}}, long before anything, except computers, were actually connected to the Internet. 

As such, from its birth, a crucial requirement for IoT, beyond ubiquitous computing, was ubiquitous connectivity. Thus, IoT considers any other object that is aware of its context and is able to communicate with other entities. Initially, RFID was the dominant technology behind IoT development, but, as of today,  wireless sensor networks (WSN) and Bluetooth-enabled devices augmented the mainstream adoption of the IoT trend~\cite{AshtonKevin2009}.

IoT has been identified as an enabler of machine-to-machine, human-to-machine, and human-with-environment interactions. Thus, IoT empowers the so-called \textbf{human-in-the-loop} systems, in which humans and things operate in a synergistic and/or cooperative fashion~\cite{Stankovic}. As new applications will intimately involve humans, a range of new opportunities to a broad range of applications including energy management, and automotive systems appear. However, several challenges arise from the human-in-the-loop, as pointed out by John Stankovic: the need for a comprehensive understanding of the complete spectrum of types of human-in-the-loop controls, need for extensions to system identification or other techniques to derive models of human behaviours and determining how to incorporate human behaviour models into the formal methodology of feedback control~\cite{Stankovic}.

A key vision of IoT is the Industrial IoT (IIoT), core technological component of the Industry 4.0 initiative. This form of IoT applications favoured by big high-tech companies envision the sensing and actuating capabilities of \textit{things} as a way to gather more data about processes, enable companies to detect and resolve problems faster, thus resulting in overall money and time savings~\cite{Buyya2016}. As an example, in a manufacturing company, IIoT can be used to efficiently track and manage the supply chain, perform quality control and assurance. 

Altogether with the recognized impact that IoT can have on the industry, it is also envisioned the impact that IoT can have on improving the quality of life~\cite{Buyya2016}. From a healthcare perspective, IoT can be a facilitator of data collecting (e.g. heart rate) which enables remote patient monitoring, \textit{viz.} ambient assisted living~\cite{Dohr2010}. Further, monitoring hazard environmental conditions can give data insights for authorities to better act and alert the population.


The IoT vision of a more connected world has been embraced by several companies and organizations. Cisco coined back in 2013 the term \textit{Internet of Everything} (IoE) as a step beyond the IoT, consisting of ``\textit{bringing together people, process, data, and things to make networked connections more relevant and valuable than ever before-turning information into actions that create new capabilities, richer experiences, and unprecedented economic opportunity for businesses, individuals, and countries}''~\cite{ciscoioe}. Shortly, Cisco concept of IoE has four pillars: \textbf{people} (connecting people in more relevant, valuable ways), \textbf{data} (convert data into intelligence to make better decisions), \textbf{processes} (delivering the right information to the right person — or machine — at the right time), and \textbf{things} (so-called IoT).

As of today, the joint technical committee of the International Organization for Standardization (ISO) and the International Electrotechnical Commission (IEC) --- ISO/IEC JTC 1 --- accepted the definition of Internet-of-Things is~\cite{isodef}:
\begin{displayquote}
	\textit{An infrastructure of interconnected objects, people, systems and information resources together with intelligent services to allow them to process information about the physical and the virtual world and react.}
\end{displayquote}

\subsubsection{Application Scenarios}

IoT has been the main booster of technological innovation in different contexts and scenarios since it works as the foundation for any kind of \textit{smart space}. Such a role is visible through the work done by the Cluster of European Research Projects on the Internet of Things (CERP-IoT) which has identified a large number of application domains for IoT~\cite{eu10}. 

\begin{table}[h]
	\centering
	\caption[IoT Application Domains]{IoT Application Domains~\cite{eu10}.}
	\label{table:domains}
	\begin{tabularx}{\textwidth}{p{2cm} X X}
		\toprule
		\textbf{Domain} & \textbf{Description}                                                                                          & \textbf{Indicative Examples}                                                                                                           \\ \midrule
		Industry        & Activities involving financial or commercial transactions between companies, organisations and other entities & Manufacturing, logistics, service sector, banking, financial governmental authorities, intermediaries, etc.                            \\ \midrule
		Environment     & Activities regarding the protection, monitoring and development of all natural resources                      & Agriculture \& breeding, recycling, environmental management services, energy management, etc.                                         \\ \midrule
		Society         & Activities/ initiatives regarding the development and inclusion of societies, cities and people               & Governmental services towards citizens and other society structures (e-participation), e-inclusion (e.g. aging, disabled people), etc. \\ \bottomrule
	\end{tabularx}
\end{table}

The CERP-IoT report~\cite{eu10} defines three IoT application domains, as they are described in Table~\ref{table:domains}. Within these application domains, several fields with open opportunities are presented such as aerospace and aviation (systems status monitoring, green operations), automotive (systems status monitoring, vehicle-to-vehicle and vehicle-to-infrastructure communication), telecommunications, intelligent buildings (automatic energy metering, home automation, wireless monitoring), healthcare (personal area networks, monitoring of parameters, positioning, real-time location systems), independent living (wellness, mobility, monitoring of an ageing population), retail, logistics, supply chain management, people and goods transportation, media, entertainment and insurance. 

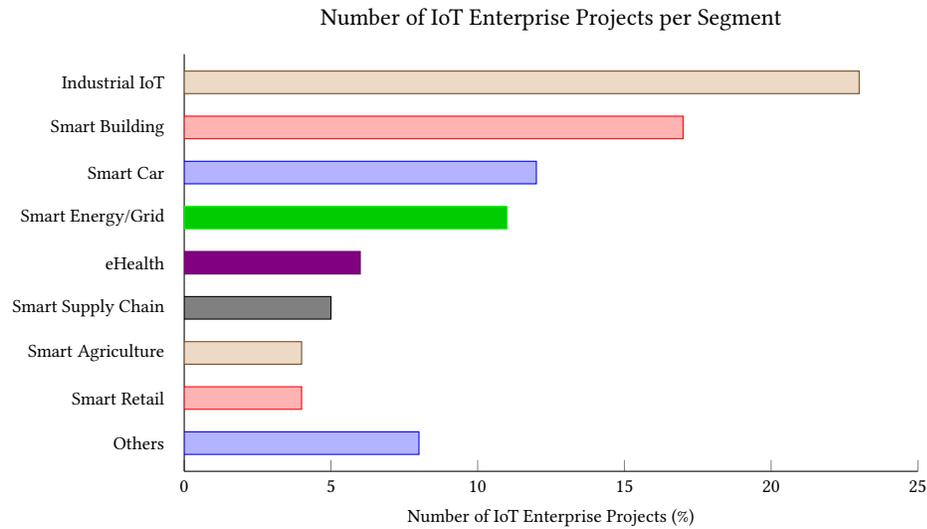
\begin{figure}[h]
	\centering
	\begin{tikzpicture}
	\begin{axis}[
	title={Number of IoT Enterprise Projects per Segment},
	xbar=0pt,
	/pgf/bar shift=0pt,
	legend style={
		legend columns=4,
		at={(xticklabel cs:0.5)},
		anchor=north,
		draw=none
	},
	ytick={0,...,8},
	ytick style={draw=none},
	axis y line*=none,
	axis x line*=bottom,
	tick label style={font=\footnotesize},
	xlabel={Number of IoT Enterprise Projects (\%)},
	legend style={font=\footnotesize},
	label style={font=\footnotesize},
	xtick={0,5,...,25},
	width=.75\textwidth,
	bar width=3mm,
	yticklabels={{Others},
		{Smart Retail},
		{Smart Agriculture},
		{Smart Supply Chain},
		{eHealth},
		{Smart Energy/Grid},
		{Smart Car},
		{Smart Building},
		{Industrial IoT},
		{Smart City}},
	xmin=0,
	xmax=25,
	area legend,
	y=6mm,
	enlarge y limits={abs=0.625},
	every axis plot/.append style={fill}
	]
	\addplot coordinates {(8,0)};
	\addplot coordinates {(4,1)};
	\addplot coordinates {(4,2)};
	\addplot coordinates {(5,3)};
	\addplot coordinates {(6,4)};
	\addplot coordinates {(11,5)};
	\addplot coordinates {(12,6)};
	\addplot coordinates {(17,7)};
	\addplot coordinates {(23,8)};
	\end{axis}  
	\end{tikzpicture}
	\caption[Statistics based upon 1600 public known enterprise IoT projects]{Statistics based upon 1600 public known enterprise IoT projects circa 2018 (not including consumer level IoT projects such as werables and smart homes)~\cite{IoTAnali}.}
	\label{fig:chartiot}
\end{figure}

The IoT Analytics GmbH report~\cite{IoTAnali} points that the most relevant enterprise-level IoT segments are Smart City, Industrial IoT, Smart Building, Smart Car, Smart Energy/Grid, eHealth, Smart Supply Chain, Smart Agriculture, Smart Retail, and their relevance is shown in the chart on Figure~\ref{fig:chartiot}. However, this report does not count with consumer level IoT segment (e.g. wearables and smart homes).

IoT enterprise applications can also be aggregated in three major categories, depending on their role, namely~\cite{Buyya2016}: \begin{inparaenum} \item monitoring and actuating, \item business process and data analysis, and \item information gathering and collaborative consumption.\end{inparaenum}

Exploiting the open IoT opportunities and different application scenarios can lead to, on one hand, improve people's quality of life, and, on the other hand, improve the industry and the enterprise world.

\subsection{Software Development Life-Cycle}

Software Development Life-Cycle (SDLC) is a process of building or maintaining software systems. Typically, it includes various phases, from preliminary development analysis (e.g. requirements, architectural design) to post-development software testing and evaluation (e.g. verification and validation)~\cite{leau2012software}. 

\begin{figure}
	\centering
	\includegraphics[width=\linewidth]{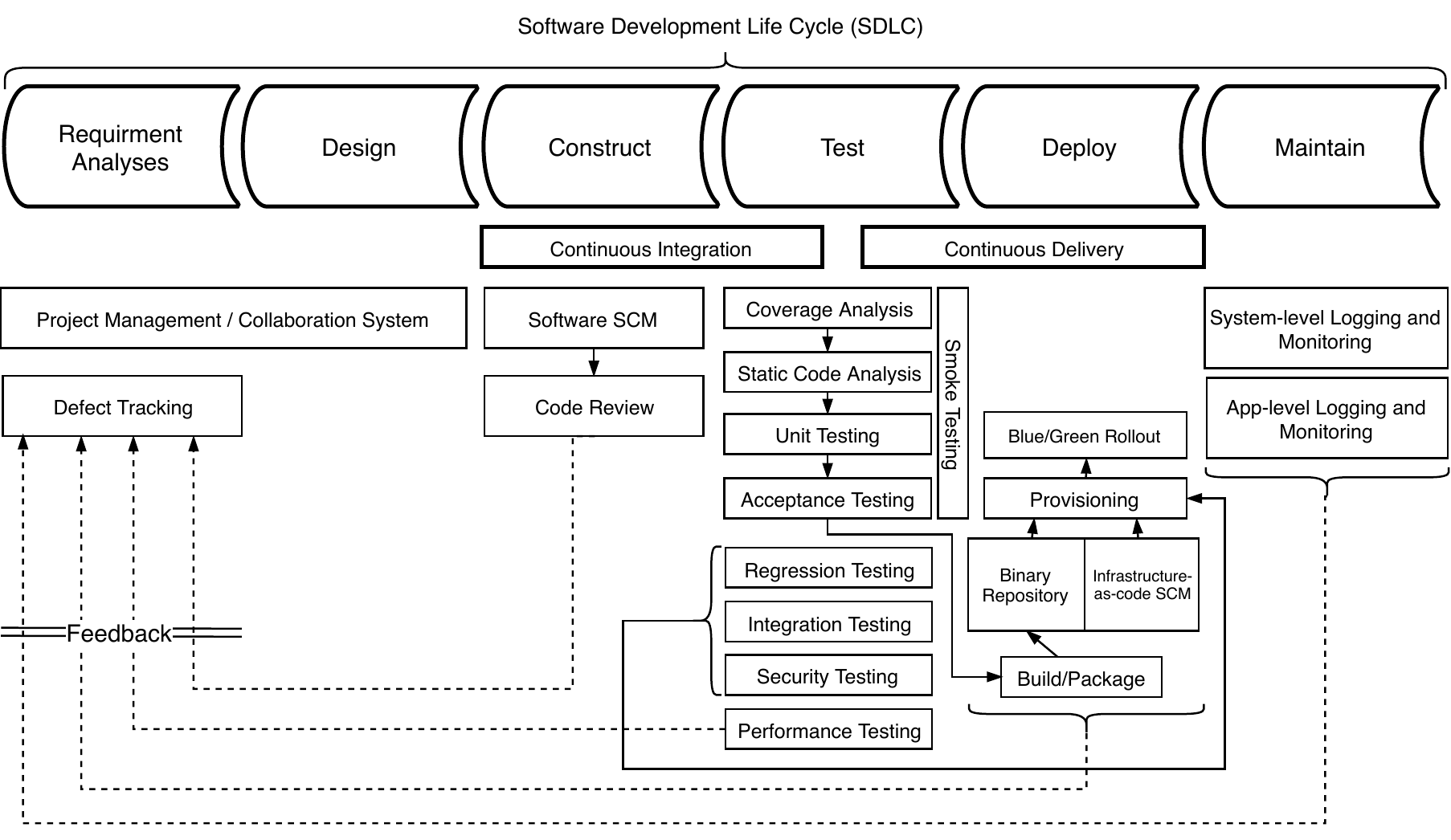}
	\caption[Software Development Life Cycle holistic view]{Software Development Life Cycle holistic view~\cite{Feeney2016}.}
	\label{fig:sdlc}
\end{figure}

SDLC also encompasses the models and methodologies that the development teams use to develop software systems, in which the methodologies form the framework for planning and controlling the entire development process. Currently, there are two SDLC methodologies categories, the Traditional Software Development ones (e.g. waterfall, Rational Unified Process) and the AGILE Software Development ones (e.g. SCRUM)~\cite{leau2012software}. A holistic view of an SDLC is depicted on Figure~\ref{fig:sdlc}. 

The application of the widespread SDLC processes to design, construct, test, deploy and maintain IoT systems faces different challenges than the ones that are faced when developing traditional software systems, due to the inherent peculiarities of the IoT ecosystem.

Primarily, from a technological viewpoint, there is a considerable amount of gaps on the software engineering body-of-knowledge regarding IoT, including, but not limited to, design, developing and testing. Secondarily, from the work team viewpoint, there is the need for the developers to have a broader base of knowledge that ranges from ultra-large-scale systems to embedded system's programming. 

\subsection{Design Patterns}

Christopher Alexander presented in his book \textsc{A Pattern Language}, circa 1977, the concept of a pattern as a way to document the architecture and urban design solutions at such time\footnote{Alexander had an M.Sc. in Mathematics and was a civil architect, and his theories on patterns emerged from the observation of cities and buildings.}. He further stands that~\cite{Alexander1977}:
\begin{displayquote}
	\textit{Each pattern describes a problem which occurs over and over again in our environment, and then describes the core of the solution to that problem, in such a way that you can use this solution a million times over, without ever doing it the same way twice.}
\end{displayquote} 

On his research, he further extended the notion of patterns beyond the <\textit{problem, forces, solution}> triplet, towards a \textit{pattern language}, which also considers the relationship between different patterns in a specific domain.

These concepts were later borrowed by the software engineering community as a way to capture and share practical knowledge and experience~\cite{Meszaros1997}. It is widely accepted that a pattern corresponds to a recurrent solution for a specific problem, that is able to achieve an optimal balance among a set of forces in a specific context, yet taking into account the consequences of it.

In software engineering the use of patterns ranges from the high-level \textit{architectural patterns}, through \textit{design patterns} until low-level \textit{idioms}~\cite{Bushmann1996}:

\begin{itemize}
	\item\textbf{Architectural patterns} express fundamental structural organisation schemes for software systems, decomposing them into subsystems, along with their responsibilities and interrelations.
	\item \textbf{Design patterns} are medium-scale tactical patterns, specified in terms of interactions between elements of object-oriented design, such as classes, relations, and objects, providing generic, prescriptive templates to be instantiated in concrete situations. They do not influence overall system structure but instead define micro-architectures of subsystems and components.
	\item \textbf{Idioms} (also known as coding patterns) are low-level patterns that describe how to implement particular aspects of components or relationships using the features of a specific programming language.
\end{itemize}

The knowledge contained in these patterns and pattern languages is a result of a synthesising process of systematic analysis and documentation of scattered empirical knowledge, and has, as of today, a profound impact in the way that developers design, build and manage software artefacts.


\section{Designing the Internet-of-Things}

\subsection{Architectural Styles}
\label{ssec:architecturestyles}

There are several basic building blocks for IoT systems, which have been around for many years, such as sensory devices, remote service invocation, communication networks and context-aware processing of events. The IoT initiative tries to leverage these well-known building blocks in a unified fashion, where the smart objects and the human beings responsible for operating them (if needed) are capable of universally and ubiquitously communicating with each other. 

A holistic system architecture for IoT needs to guarantee flawless operation of its components and \textit{fuse} the physical and virtual realms. For reaching such an objective, the IoT systems need to be dependable, adaptable, handle dynamic interactions, highly-scalable and human-centric~\cite{Buyya2016}.

These systems follow an architectural style that is mostly compatible with nowadays standards, both in communication aspects and design aspects. Thus, the most common foundation of IoT systems is Web Services, influencing the way they are built and communicate.

As such, IoT systems are usually based on either a Representational State Transfer (REST\footnote{Also known as RESTful, which is used to describe services that follow a REST architecture.}) architecture or a Simple Object Access Protocol (SOAP) architecture. Nonetheless, either architecture is service-oriented (SOA), providing a set of services by exposing their own arbitrary sets of operations, allowing interoperability among the heterogeneous devices~\cite{Buyya2016}. 

On one hand, SOAP is a more traditional architectural style, being heavier in terms of bandwidth, more complex and use Extensible Markup Language (XML) data-exchange format. On the other hand, REST is more flexible, coupled with JavaScript Object Notation (JSON) data-exchange format, and is generally faster and uses less bandwidth.

There are two main architectural approaches when developing IoT systems, namely, mashup-based and model-based~\cite{Prehofer2013,Prehofer2015}. In mashup-based approaches, systems are developed by composing, or mashing up, existing services. Thus, mashups are often used for personalized, situational, short-lived and non-business critical applications developed, typically based upon familiar web development tools and technologies (e.g. application prototyping)~\cite{Blackstock2012}. Model-based approaches base itself on the ability to describe a system on a higher level of abstraction, thus permitting a very expressive modelling of systems, possibly with code generation~\cite{Prehofer2013,Prehofer2015}. Combinations of this two approaches (a hybrid between mashup and model-based) are not found in the literature.

\subsection{IoT Interoperability Standards}

Different entities have been working on different standards to ensure an interoperable Internet-of-Things, simply put a common language that devices can speak between them and different applications, thus, reducing the IoT fragmentation. A summary of the most known initiatives is given on Table~\ref{table:iotapi}, and the most widespread are analysed in the following paragraphs.

\begin{table}[h]
	\centering
	\caption{Overview of the IoT Enabling Models and API's}
	\label{table:iotapi}
	\begin{tabularx}{\textwidth}{P{3.3cm} X P{2.1cm}}
		\hline
		\textbf{Name} & \textbf{Description}                                                                                                                                              & \textbf{Status}                                                              \\ \hline
		Web Thing API~\cite{webmoz}             & Common data model and API for the Web of Things.                                                                                                                  & Last Draft May / 2018                           \\ \hline
		OGC SensorThings API~\cite{ogciot}      & An open, geospatial-enabled and unified way to interconnect IoT devices, data, and applications over the Web.                                                     & Version 1.0 July / 2016      \\ \hline
		IOTDB~\cite{iotdbs}                    & A semantic layer for the IoT, which includes definitions for all the data, to provide both,formal definitions for all important items and unlimited expandability. & Last Draft April / 2017                                       \\ \hline
		Web Thing Model~\cite{webthingw3c}           & A common model to describe the virtual counterpart of physical objects in the Web of Things.                                                                      & Last Submission August 2017                   \\ \hline
		SENML~\cite{SENML2018} &  The Media Types for Sensor Markup Language is a standard for representing simple sensor measurements and device parameters.  &     Last Version April / 2013     \\ \hline
		LsDL~\cite{Ray2017}  &   The \textit{Lemonbeat smart Device Language} is a XML-based smart devices encoding language that is read as Lemonbeat smart Device Language (LsDL). &      Not Available    \\ \hline
	\end{tabularx}
\end{table}

\begin{description}
	\item[Web Thing API by Mozzilla~\cite{webmoz}] Mozilla proposal of a common data model and API for the Web of Things. The Web Thing Description provides a vocabulary for describing physical devices connected to the World Wide Web in a machine-readable format with a default JSON encoding. The Web Thing REST API and Web Thing WebSocket API allow a web client to access the properties of devices, request the execution of actions and subscribe to events representing a change in state. Some basic Web Thing Types are provided and additional types can be defined using semantic extensions with JSON-LD. Also, the proposal includes details on Web of Things Gateway Protocol Bindings which proposes non-normative bindings of the Web Thing API to various existing IoT protocols and a set of Web of Things Integration Patterns which provides advice on different design patterns for integrating connected devices with the Web of Things, and where each pattern is most appropriate.
	
	\item[OGC SensorThings API by Open Geospatial Consortium (OGC)~\cite{ogciot}] An open, geospatial-enabled and unified way to interconnect the Internet of Things (IoT) devices, data, and applications over the Web. At a high level, the OGC SensorThings API provides two main functionalities and each function is handled by a part. The two parts are the Sensing part and the Tasking part. The Sensing part provides a standard way to manage and retrieve observations and metadata from heterogeneous IoT sensor systems.  The Tasking part is still under development.
	
	\item[IOTDB~\cite{iotdbs}] Things are described by semantically annotating the \textit{data associated with the Thing}, being this description built from \textit{composition} of \textit{atomic} elements (that cannot be meaningfully subdivided further) and are \textit{extensible} (allowing to add in elements from other Semantic ontologies). The Things can have many different \textit{bands} of data associated with them (e.g. the metadata, the actual state). 
	
	\item[Web Thing Model by W3C~\cite{webthingw3c}] A common model to describe the virtual counterpart of physical objects in the Web of Things. It defines a model and Web API for Things to be followed by anyone wanting to create a product, device, service, or application for the Web of Things. The model and protocols proposed aim at making the interaction between Things in the Internet of Things accessible through Web standards to facilitate the implementation of Web applications making use or retrieving data from real-world objects. 
\end{description}

\subsection{Internet-of-Things Patterns}
\label{sec:designpatternsiot}

Patterns have been used for long to document recurrent solutions to common problems across different areas of software engineering and others. A similar effort has been carried out in the Internet-of-Things field trying to document the common solutions to the complexity and common challenges of developing these systems. Reinfurt et al.~\cite{Reinfurt2016,Reinfurt2017} works are of the first ones in this area and the most relevant patterns already identified are summarily described below:

\begin{description}
\item[Device Gateway]
\textit{Some devices cannot directly connect to a network because they do not support the required communication technologies. These devices can be connected through a translating gateway~\cite{Reinfurt2016}.}

Due to the high heterogeneity of devices and network protocols that are part of the IoT ecosystem, there are several limitations when making these devices interoperate between themselves and the surrounding services and infrastructures. A common solution is the use of \textit{device gateways} that overcome the limitations of interoperability by translating the data to the standard being used (e.g. JSON).

\item[Device Shadow]
\textit{Other components can interact with currently offline devices by communicating with a persistently stored virtual representation of the device that is synchronised once the device reconnects~\cite{Reinfurt2016}.}

Dealing with a large-scale and highly-distributed, geographically and in terms of processing, jointly with high-latency and low-reliability networks, results in devices that sometimes are not available. Thus, the need appears of having a virtual representation of the devices (avatar) that mocks the same behaviour of the actual device even when the device is offline. If the device comes online the \textit{avatar} is responsible for the synchronisation of the device state.

\item[Rules Engine]
\textit{Users can define simple rules without needing to program. These rules tell the system with what action it should react to incoming events~\cite{Reinfurt2016}.}

As a way of abstracting low-level IoT interactions a typical solution is to provide the user a way to simply define \textit{if-this-then-that} rules, that allow them to add simple and personalised logic to their devices and services.

\item[Device Wakeup Trigger]
\textit{A device that is not currently connected to the backend server can be informed to do so by sending a message to a low-power communication channel where the device listens for such messages~\cite{Reinfurt2016,Reinfurt2017}.}

There are cases when low-powered devices enter a power-saving mode (sleep-mode) and only wake up and reconnect to the network from time to time in order to transmit new data. However, there are scenarios where the devices need to be connected during their sleep period. Thus, mechanisms must be put in place in order to be able to send a trigger message to the device via a low energy communication channel.

\item[Remote Device Management]
\textit{When dealing with numerous devices remotely set up a management service and deploy a management agent to all the devices that need to be controlled~\cite{Reinfurt2017}.}

When dealing with a high number of remote devices, a management service should be set up, along with the deploy of managing agents in the devices. This allows the operator to remotely manage the devices by the means of commands that are then run on the devices by the means of the deployed agent.

\item[Remote Lock and Wipe]
\textit{When a device is lost or stolen, its functionality can be remotely locked or data on it can be wiped, either fully or partially, to protect it from possible  attacks~\cite{Reinfurt2016,Reinfurt2017}.}

There are situations where devices are in situations that are easily lost or stolen. In order to prevent an attacker from misusing the functionality of the device (read stored data or gain access to the network) the device must be able to receive instruction from an authority in order to delete files, folders, applications or memory areas, revoke or remove permissions, keys, and certificates.

\item[Delta Update]
\textit{To reduce the size of messages containing sensor data without losing any information the last message sent should be stored and the delta from the current data to this message calculated, and only send the delta to the receiver~\cite{Reinfurt2017}.} 

There are situations where devices produce large amounts of data by repeating non-changing values over and over again (e.g. humidity sensor). In order to reduce the data throughput of these devices, the last message sent should be always stored, and a delta between the new data and the last sent message calculated. Then, the only data that need to be sent over the network is the delta value and a hash of the data, and the receiver side can calculate the resulting data and check the integrity of the result.

\item[Visible Light Communication]
\textit{In a situation where wireless communication is not viable use visible light for short-range  wireless communication~\cite{Reinfurt2017}.}

In some situations, when the radio spectrum is not a viable form of communication, several experiments have been carried out on the use of visible light communication for short-range data transmission.

\item[Device Registry]
\textit{In a large-scale and dynamic topology environment, it is not viable to use a local registry in each device, thus it is desirable to have a module responsible to store and answer for queries of devices endpoints~\cite{Ramadas2017}.}

\item[Device Raw Data Collector]
\textit{Both the raw data collected by the devices and the logs they produce can be a source of information about the wealth of the system and devices, thus this data should be collected and stored~\cite{Ramadas2017}.}

\item[Device Error Data Supervisor] 
\textit{Due to the high-complexity of IoT systems, both in number and heterogeneity of devices, occurrence of errors in edge devices is more common. A supervisor of the logs produced by the system entities can handle and process errors, enabling the mitigation of failures~\cite{Ramadas2017}.}

\item[Predictive Device Monitor] 
\textit{T high-complexity of IoT systems, both in number and heterogeneity of devices, increase the occurrence of errors in edge devices. A predictive-capable module can watch the logs produced by the system entities predict device malfunctions, thus enabling the prevention of failures~\cite{Ramadas2017}.}

\end{description}

\subsection{Implications and Barriers}

\subsubsection*{Resource Management}

IoT can be pictured as a big graph, with numerous nodes with different resource capacity (computing, storage, connectivity). As a consequence, the selection and provisioning of such resources have a great impact on the Quality of Service (QoS) of the IoT applications~\cite{Buyya2016}.

Considering a large-scale scenario such as a smart-city, efficient resource management becomes a priority due to the need of robustness, fault-tolerance, scalability, energy efficiency, QoS, and service-level agreements (SLA)~\cite{Buyya2016}.

\subsubsection*{Identification and Resource/Service Discovery}

Discovery in IoT systems is twofold. The first objective is to identify and locate the \textit{things} in the system,  which can be achieved by storing and indexing metadata information about each device, and the second objective is to discover the target service that needs to be invoked for a given task~\cite{Buyya2016}.

In order to fulfil such need, an effective discovery mechanism is necessary. Such a mechanism is required to have considerations in order to minimize the consumed energy, latency and impact on the final user experience. However, the heterogeneous nature of the devices, variable data types, concurrent operations and the confluence of data from devices exacerbates this task~\cite{Gubbi2013}.

\subsubsection*{Identity Management and Authentication}

To identify the billions of connected devices that make part of the IoT vision, setting their access level on the whole ecosystem is a must. Thus, IoT devices have to be uniquely identified~\cite{Buyya2016}. Solutions such as \textit{ucode}\footnote{\textit{ucode} generates 128-bit codes that can be used in active and passive RFID tags.} and Electric Product Code (EPC)\footnote{EPC creates unique identifiers using Uniform Resource Identifier (URI) codes.} reduce the complexity of expanding the local environment and linking it with larger ecosystems.

However, identity requirement is not yet adequately met in networks, with only a few solutions have been proposed regarding this issue~\cite{Sicari2015}. Furthermore, Sicari et al. raise some extra questions regarding the identity problem and access control perspective~\cite{Sicari2015}:

\begin{itemize}
	\item \textit{To manage access control, how could the IoT system deal with the registration of users and things and the consequent issuance of credentials or certificates by authorities?}
	\item \textit{Could the users/things present these credentials/certificates to the IoT system in order to be allowed to interact with the other authorized devices?}
	\item \textit{Could a following step be the definition of specific roles and functions within the IoT context, in order to manage the authorization processes?}
\end{itemize}

Few solutions have been proposed in order to address these questions, and the use of a subscriber method with a group membership scheme to deal with the access control of heterogeneous devices has been proposed as one possible solution~\cite{Sicari2015}.

Further, from an authentication perspective, some solutions exist for constrained devices, such as the proposed by Sicari et al. that combines Physical Unclonable Functions (PUF) with Embedded Subscriber Identity Module (eSIM). Their solution provides cheap, secure, tamper-proof secret keys to authenticate constrained M2M devices while guaranteeing scalability, interoperability, and compliance with security protocols~\cite{Sicari2015}.

\subsubsection*{Data Management and Analytics}

\begin{figure}[h]
	\centering
	\includegraphics[width=0.75\linewidth]{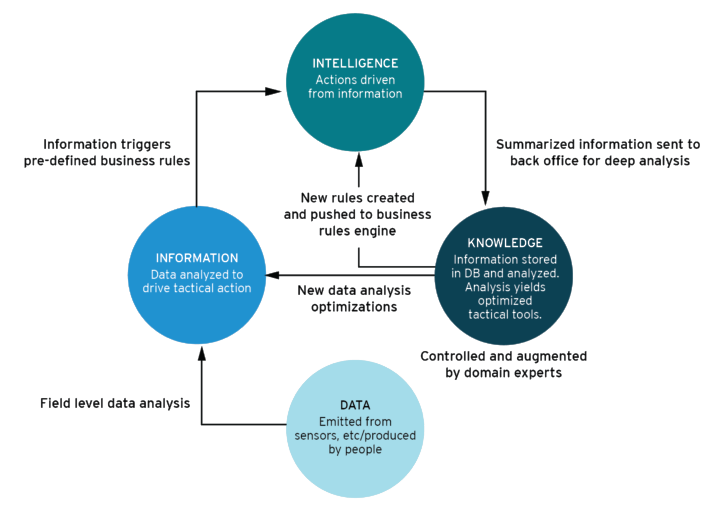}
	\caption[IoT information life cyle]{RedHat vision of the life cycle of IoT originated information, from collection to supported decision making~\cite{James15}.}
	\label{fig:iot-information}
\end{figure}

IoT brought a new vision on collecting data about systems and surroundings, giving more sensing possibilities to almost any situation. As a consequence, IoT has become one of the biggest sources of data nowadays, for both individuals and organizations. It is commonly accepted that the real power of IoT resides on this data collection and analysis, as depicted on Figure~\ref{fig:iot-information}\cite{James15}.

The amount of data generated by some of these sensing networks fit on the view of Big Data, since IoT data, like Big Data, is characterized by 3Vs, namely velocity, volume, and variety. So, as a Big Data system, IoT shares the same needs and challenges of a typical Big Data scenario. Thus, the volume, velocity, and variety (not to mention variable veracity) make the storing and analytics approach that will generate useful insights, a very complex one, leading to a possible data overload (too much data without value or inability to analyse the data)~\cite{RisteskaStojkoska2017}. As such, several open research questions appear for IoT in the context of Big Data (e.g. traditional SQL-queried relational database management systems (RDBMSs) are unsuitable for IoT needs).

The problem is even more complex when factors such as data integrity are taken into account, not only because of their impact on the quality of service but also for its security and privacy related aspects especially on outsourced data~\cite{RisteskaStojkoska2017}.

Yet, several advancements of the Big Data research appear as the solution for IoT needs, such as lambda architectures, stream processing, batch processing, and time-series oriented databases~\cite{Buyya2016}.

But finally, as Hurlburt et al. point, a more fundamental question rises~\cite{Hurlburt2012}:

\begin{displayquote}
	\textit{As the IoT becomes ubiquitous, issues of information ownership will become crucial. Who will own the oceans of data IoT will generate?}
\end{displayquote}

Further, taking into consideration the Open Data\footnote{Open Data is the idea that some data (and metadata) should be freely available to everyone to use and republish as they wish, without restrictions from copyright, patents or other mechanisms of control.} momentum, questions arise if anyone should own the data generated at any instance, especially for government-based IoT scenarios (e.g. Smart Cities).

\subsubsection*{Security and Privacy}

The spreading of IoT usage increased the size of the attack surfaces that should be taken into account by manufacturers, developers, security researchers and those looking to deploy or implement new IoT applications. Sicari et al. survey point out that there are eight main categories of security concerns that must be considered in the IoT landscape, namely: authentication, access control, confidentiality, privacy, trust, secure middleware, mobile security, and policy enforcement~\cite{Sicari2015}.  

IoT devices typically have resource constraints thus the implementation of standard security mechanisms is not feasible (e.g. cryptographic algorithms need considerable bandwidth and energy to provide end-to-end protection). As a consequence, the wireless communication used by the majority of these devices are vulnerable to eavesdropping and man-in-the-middle attacks.

There are already a considerable number of occasions where the leverage by malicious parties of the lack of a security layer on top of IoT application leads to nefarious consequences. 

\begin{figure}[h]
	\centering
	\includegraphics[width=0.95\linewidth]{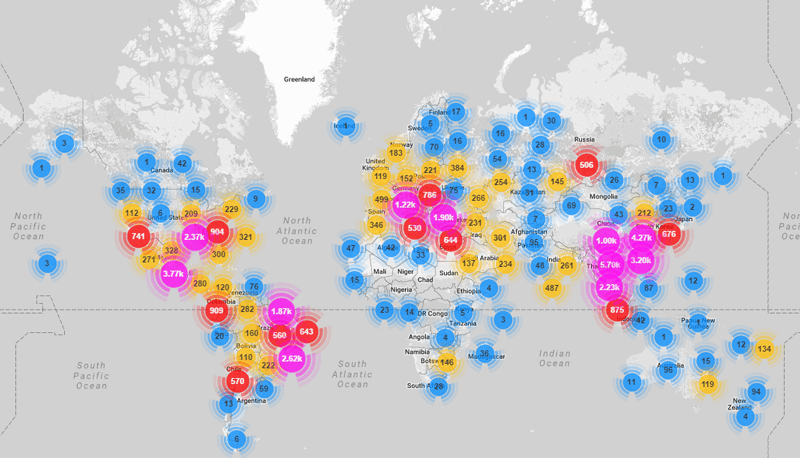}
	\caption[Geo-location of Mirai-infected devices]{Geo-location of Mirai-infected devices in the world. The study carried by Imperva pointed that above 49 thousand of IoT devices were infected by the Mirai malware~\cite{Herzberg2016Oct}.}
	\label{fig:mirai-botnet-map}
\end{figure}

One of the most recent events that have resulted from the lack of security in the IoT is the case of Mirai botnet~\cite{Sicari2015}. The Mirai malware took advantage of misconfiguration of IoT devices (default passwords of the telnet or SSH accounts) in order to gain shell access. Mirai has responsibility for the vast Internet outage in October 2016 which was caused by the various DDoS attack using above 49 thousand infected IoT devices around the world (Figure~\ref{fig:mirai-botnet-map})~\cite{Dias2017}.

The Open Web Application Security Project (OWASP) has been working towards a reference documentation on how to tackle the IoT security perspective. As described, this project \textit{``is designed to help manufacturers, developers, and consumers better understand the security issues associated with the Internet of Things, and to enable users in any context to make better security decisions when building, deploying, or assessing IoT technology``}~\cite{OWASP2018Apr}.

From a privacy perspective, IoT introduces a whole new degree of concerns for consumers. These concerns are not only because of the ability of these devices to collect personal information like users' names and telephone numbers but because these devices can also monitor user activities (e.g., when users are in their houses and what they had for lunch)~\cite{Sicari2015}. Standardization and regulatory (legal) limitations and gaps are a key problem in this field~\cite{Buyya2016}.

\subsection{Cloud, Fog, and Mist/Edge Computing}

Given the high data volume being generated by IoT, plus the variety of objects that make part of it, there are several issues/requirements that need to be addressed in order to provide a good QoS, such as minimizing the latency (milliseconds matter for many types of industrial systems, such as when you are trying to prevent manufacturing line shutdowns or restore electrical service), conserving network bandwidth (since it is not practical to transport vast amounts of data from thousands of edge devices to the cloud) and increasing local efficiency (e.g. collecting and securing data across a wide geographic area with different environmental conditions may not be useful)~\cite{book:1689191}. 

As such, traditional IT cloud computing models (direct connection between end-devices and the cloud) does not suffice the need of IoT systems, due to issues such as limitations in bandwidth in last-mile IoT networks, very high latency, network unreliability and increasing volume of the data being generated, transmitted and, \textit{a posteriori}, analysed ~\cite{book:1689191}.

\begin{figure}[h]
	\centering
	\includegraphics[width=.9\textwidth]{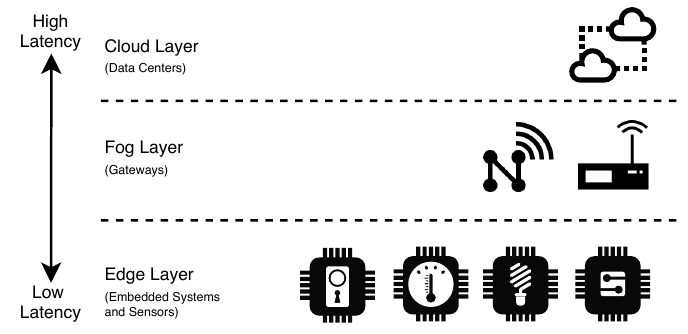}
	\caption[Common architectural layers of IoT systems]{Typical architectural layers composition  of an Internet-of-Things system. The upper layers have more latency and more computational power than the lower layers of the stack.}\label{fig:iotlayers}
\end{figure}

Fog Computing has come as one solution to the above-mentioned challenges. The concept coined by Flavio Bonomi and Rodolfo Milito of Cisco Systems circa 2015 focus on distributing data management throughout the IoT system, as close to the edge of the network as possible~\cite{book:1689191}. The National Institute of Standards and Technology (NIST)~\cite{iorga2018fog} states that the fog computing model ``facilitates
the deployment of distributed, latency-aware applications and services, and consists of  \textit{fog nodes} residing between smart end-devices and centralized (cloud) services''.

\begin{figure}[h]
	\centering
	\includegraphics[width=0.8\linewidth]{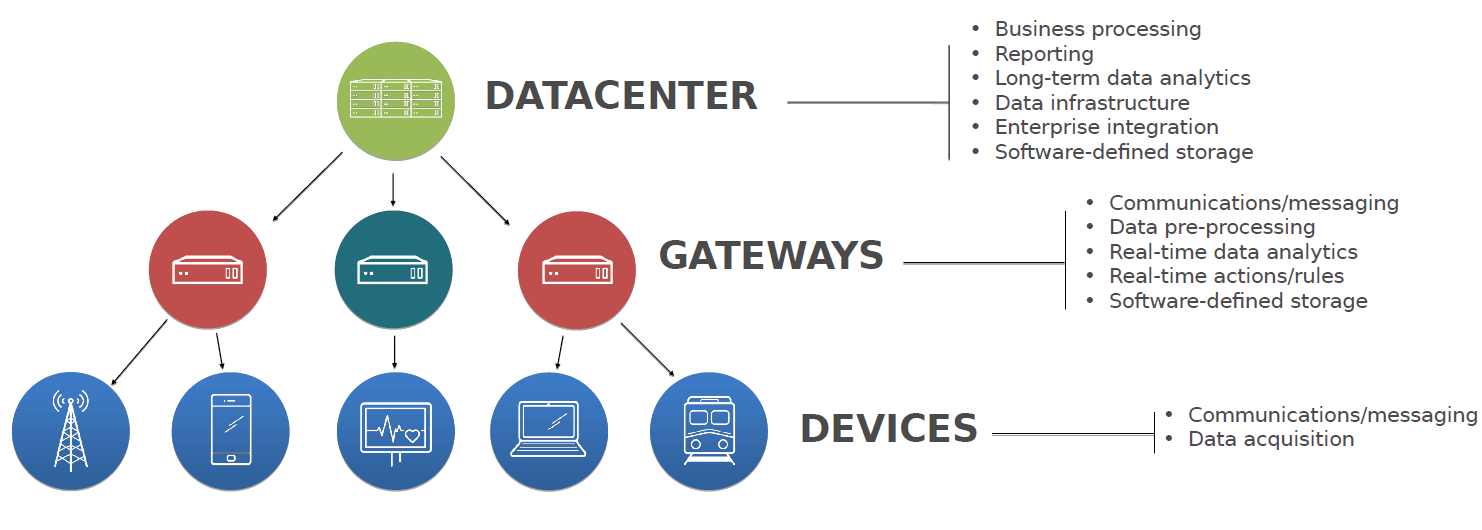}
	\caption[RedHat vision of the IoT enterprise architecture]{RedHat vision of the IoT enterprise architecture~\cite{redhatiotenter}. Generalizing, the architecture fits perfectly on the depict on Figure~\ref{fig:iotlayers}, being the edge layer the \textit{devices}, the fog layer the \textit{gateways} and the cloud layer the \textit{datacenter}.}
	\label{fig:fogcloud}
\end{figure}

As such, fog computing provides contextual location awareness, low latency, geographical distribution, heterogeneity support, interoperability, real-time interactions support, scalability and agility of federated, \textit{fog-node} clusters~\cite{iorga2018fog,book:1689191}.

Edge Computing consists of the IoT devices and sensors that often have constrained resources. However, as computing capabilities increase these devices have enough computing capabilities to perform at least low-level analytics and filtering to make basic decisions~\cite{book:1689191}. It is important to note that in some cases Edge Computing is called Mist Computing.

With the birth of Fog Computing and Edge Computing, alongside with the Cloud Computing, a new architectural hierarchy for IoT systems was born, as depicted on Figure~\ref{fig:iotlayers}. In the lower level are the IoT devices (i.e. embedded systems and sensors/actuators), the \textit{edge layer}. After, and close to the edge-devices are the \textit{fog nodes} which together create the \textit{fog layer}. On the top layer are the data centres, the \textit{cloud layer}. The applicability of this architectural hierarchy is visible in enterprise solutions such as the RedHat IoT enterprise architecture depict on Figure~\ref{fig:fogcloud}.

\section{Conclusions}



The use design patterns as a way of documenting well-stated solutions for recurring problems in the IoT domain is of great impact on reducing the problems that developers have when they have to develop such kind of systems. Although the remarkable work by Reinfurt et al., the number of patterns identified is still residual when taking into account the number of open technological challenges. The patterns are spread among architectural layers and from hardware to software perspectives. A more extensive work must be pursued on the systematization of existent solutions (in both academia and enterprise-grade solutions).

It is considered that works such as \textsc{Microservices Patterns} by Chris Richardson~\cite{microservicespatterns}, \textsc{Patterns for Fault Tolerant Software} by Robert Hanmer~\cite{hanmer2013patterns}, \textsc{Patterns for Software Orchestration on the Cloud} by Boltd et al.~\cite{sousa2015patterns} are of relevance and should be considered when delving into the patterns for IoT systems.



Furthermore, efforts towards one (or more) pattern language for IoT must be pursued, establishing relations between the founded patterns.


\section{Developing the Internet-of-Things}

\subsection{Mashup-based \& Model-based IoT Development}

\subsubsection{ThingML}
\label{sec:thingml}

ThingML (Internet of Things Modelling Languages) resulted from the work by Fleurey et al. in the scope of the HEADS EU FP7 research project and aims to transfer the promises of academic model-based software development to the industry~\cite{Harrand:2016:TLC:2976767.2976812,7958482}. It consists of an MDSE approach to tackle the complexity of developing Internet-of-Things systems, giving a tool-chain that target resource-constrained embedded systems such as low-power sensor and microcontroller-based devices. 

\begin{figure}[ht]
	\centering
	\includegraphics[width=0.75\linewidth]{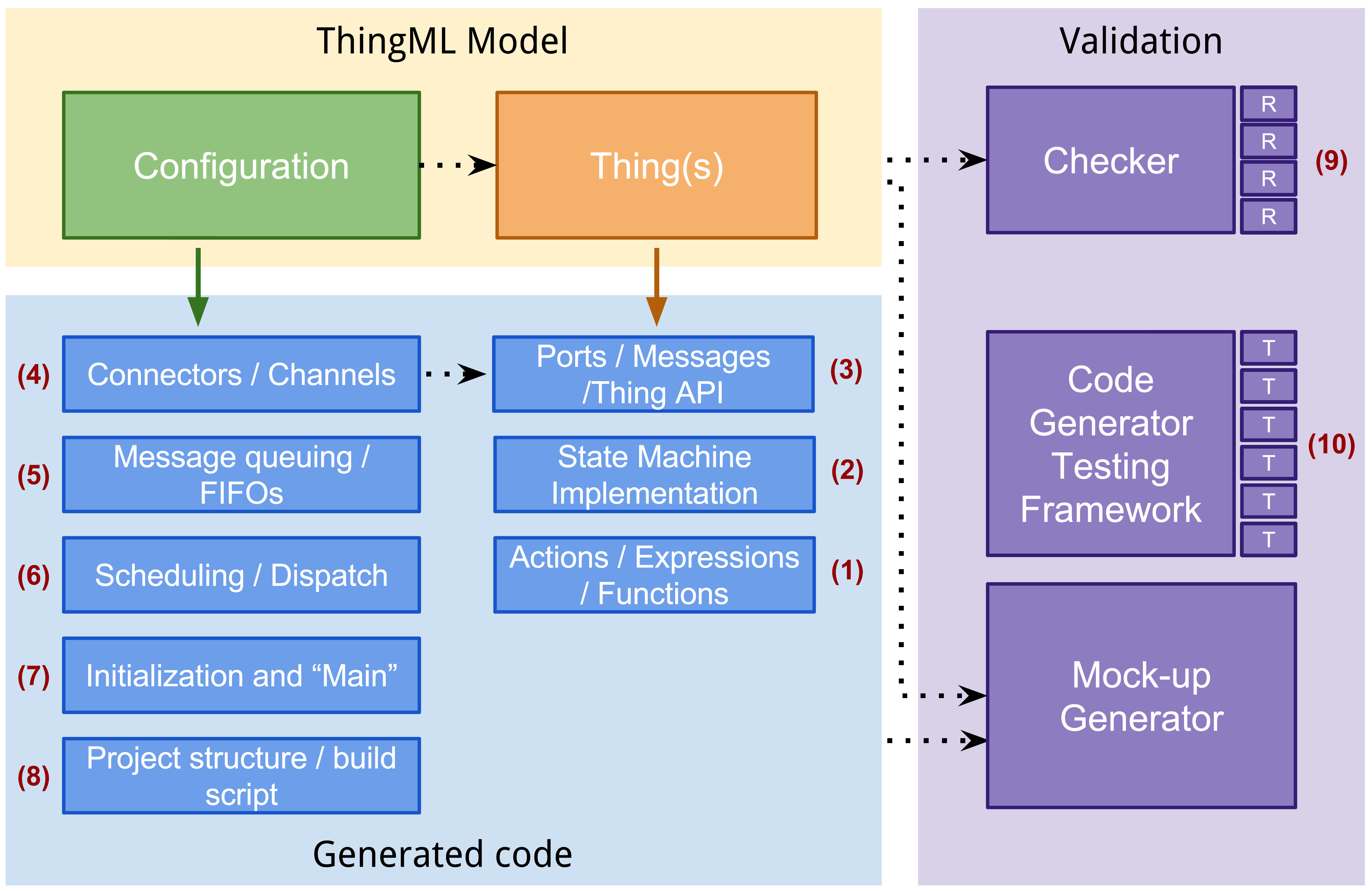}
	\caption[ThingML code generation framework]{The ThingML code generation framework. The 10 variation points of the framework are identified and separated in two groups: the ones responsible for the generation of code for \textit{things} and the ones corresponding to the generation of code for the \textit{applications} (Configuration)~\cite{Harrand:2016:TLC:2976767.2976812}.}
	\label{fig:thigmlframe}
\end{figure}

It uses a domain-specific modelling language (DSML) that allows the description of both software components and communication protocols, resulting of a combination of architecture models, state machines and an imperative action language (that allows to seamlessly interleave platforms specific code and platform independent code). Further, ThingML provides a customizable code generation framework which can be tailored to specific target languages, middleware, operating systems, libraries and even builds systems, as depicted on Figure~\ref{fig:thigmlframe}.

Summarily, the ThingML approach is composed of: 
\begin{description}
	\item[Modeling Language] The language combines a set of well-proven software modelling constructs for the design and implementation of distributed reactive systems, which includes state-charts, an imperative platform-independent action language and specific constructs targeted at IoT applications.
	\item[Toolset] In order to support the development of systems using the ThingML language, a set of tools is provided with it:
	\begin{itemize}
		\item A language editor (text editor);
		\item A set of transformations to create diagrams from ThingML models (e.g. export to UML);
		\item A multi-platform code generation framework (Figure~\ref{fig:thigmlframe}), supporting multiple programming languages (C, Java, Javascript).
	\end{itemize}
	\item[Methodology] A methodology for extending the base ThingML solution, developing systems using it and operate those systems is extensively documented and openly available.
\end{description}

The first goal of ThingML is to simplify the development of highly-distributed and heterogeneous systems by abstracting from the heterogeneous platforms and devices to model the desired IoT system's architecture. 

Focusing on the DSML, it provides a solution for both integrating off-the-shelf or legacy components as black-boxes and to model the complete behaviour of components. The original motivation by Fleurey et al. for the creation of this new DSML was that no existing modelling language provide the exact set of concepts needed, and the lack of support for all subsequent phases of the IoT life-cycle with practical tools based on the same concepts and the same well-defined semantics~\cite{7958482}. More extensively, the ThingML language features are~\cite{Harrand:2016:TLC:2976767.2976812}:

\begin{itemize}
	\item \textit{Component types with ports and asynchronous messaging}: All parts of the system need to be described as components with an asynchronous messaging interface.
	\item \textit{Composite State Machines}: The component's behaviour can be specified as \textbf{state machines}. The ThingML state machines are aligned with UML2 state charts and include \textbf{composite states}, \textbf{regions} and \textbf{history states}.
	\begin{itemize}
		\item Due to the highly-dynamical topology of IoT systems (with devices constantly coming and going in the network), ThingML state machines employ sessions similarly to user sessions in web applications. In ThingML a \textbf{session} is a dynamically instantiated parallel region, initialized with a copy of the context (set of properties) of its parent, at fork time, executes its own behaviour, communicate only through asynchronous messages and terminates when it reaches a final state.
	\end{itemize}
	\item \textit{Event-based reactive programming}: The behaviour of the different components can be expressed using event processing rules, from both Event-Condition-Action (ECA) rules to \textbf{Complex Event Processing} (CEP~\footnote{Event processing that combines data from multiple sources to infer events or patterns that suggest more complicated circumstances.}).
	\begin{itemize}
		\item The ThingML CEP works similarly to state machines (processes a set of input messages and produces output messages), however, it is fully declarative. It includes operators to join and merge streams of messages and to process messages over windows, defined by time or the number of messages~\cite{Morin2017}.
	\end{itemize}
	\item \textit{Imperative action language}: Allow the fully modelling of all conditions and actions within event processing rules plus state machines in a platform-independent way. It also includes a template language for easily embedding or linking platform-specific code, allowing to arbitrarily blending model actions with target-language actions and easily sharing variables and implementing calls and callbacks in both directions.
\end{itemize}

One key difference between UML and ThingML is that the ThingML language primary concrete syntax is lexical and not graphical. However, visual notations are the most common approach for MDE~\cite{Ferreira}. 

ThingML lacks in covering the full development life-cycle, due to limitations on software deployment and updates. Further, the ability to share computational resources and devices among IoT applications, in a reliable and foreseeable fashion, is not covered.

Finally, the concept of models@runtime is just preliminarily explored, by using ThingML models and the Kevoree system for integrating mediators as a mean to have reflection~\footnote{Kevoree provides a set of tools for creating and managing distributed systems.}~\cite{Hao2012}, and it does not exploit the potential of Live Programming.

\subsubsection{Node-RED}
\label{sec:nodered}

\begin{figure}[h]
	\centering
	\includegraphics[width=0.9\textwidth]{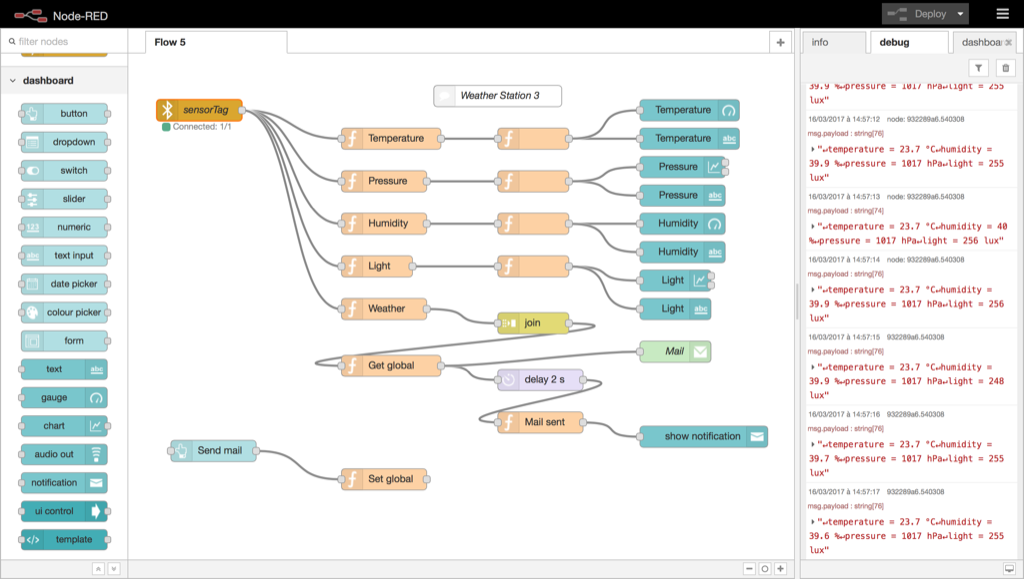}
	\caption[Node-RED development tool with example \textit{flow} and \textit{nodes}]{Node-RED development tool with example \textit{flow} and \textit{nodes}.}
	\label{fig:noderedide}
\end{figure}

Node-RED~\footnote{Node-RED, Flow-based programming for the Internet of Things, \url{http://nodered.org}} is an open-source mashup-based approach for developing Internet-of-Things systems~\cite{Blackstock2012}, originally developed by IBM Emerging Technologies and now a JS Foundation Project~\cite{lewis_2016}. The development tool is web-based, platform-agnostic, and allows developers to wire together hardware devices, APIs and online services using a visual flow-based programming model and a drag-and-drop interface, as shown in Figure~\ref{fig:noderedide}

The Node-RED runtime is built on Node.js (JavaScript), taking advantage of its built-in event model. Node-RED data-flow programs are known as \textit{flows}, consisting of \textit{nodes} connected by wires. Several \textit{node} default templates are provided that can be drag-and-drop into a flow canvas. Once the developer creates or updates a flow, it must be \textit{deployed}, both saving it to the server and (re)starting its execution~\cite{Blackstock2014}. 

Node-RED portfolio of available \textit{nodes} can be extended, by developing new nodes in JavaScript that extends the \texttt{Node} base class. This \texttt{Node} class is a subclass of an \texttt{Event Emitter} (part of the Node.js event API), that implements the observer pattern in order to maintain subscriber lists defined by the \textit{wires} emitting events to downstream \textit{nodes}~\cite{Blackstock2014}.

Input \textit{nodes} can, on instantiation,
\begin{inparaenum}
	\item subscribe to external services,
	\item listen for data on a specific port or
	\item start processing HTTP requests.
\end{inparaenum}

Once the data is processed by a given \textit{node}, either from an external service or from an upstream \textit{node}, a method is called with the resulting JSON that sends the object to downstream \textit{nodes} that can either generate additional events or push the resulting data to outside services or systems~\cite{Blackstock2014}. 

Node-RED provides an export mechanism for deploying the same flow on different Node-RED deployments, using a JSON with all the configurations.  

A modification to the original version of Node-RED has been purposed by Blackstock et al.~\cite{Blackstock2014}, in order to make it suitable for execution on a range of runtime environments in order to be able of distributing Node-RED \textit{flows}, between different servers, gateways, and devices.

\begin{figure}[h]
	\centering
	\includegraphics[width=0.85\textwidth]{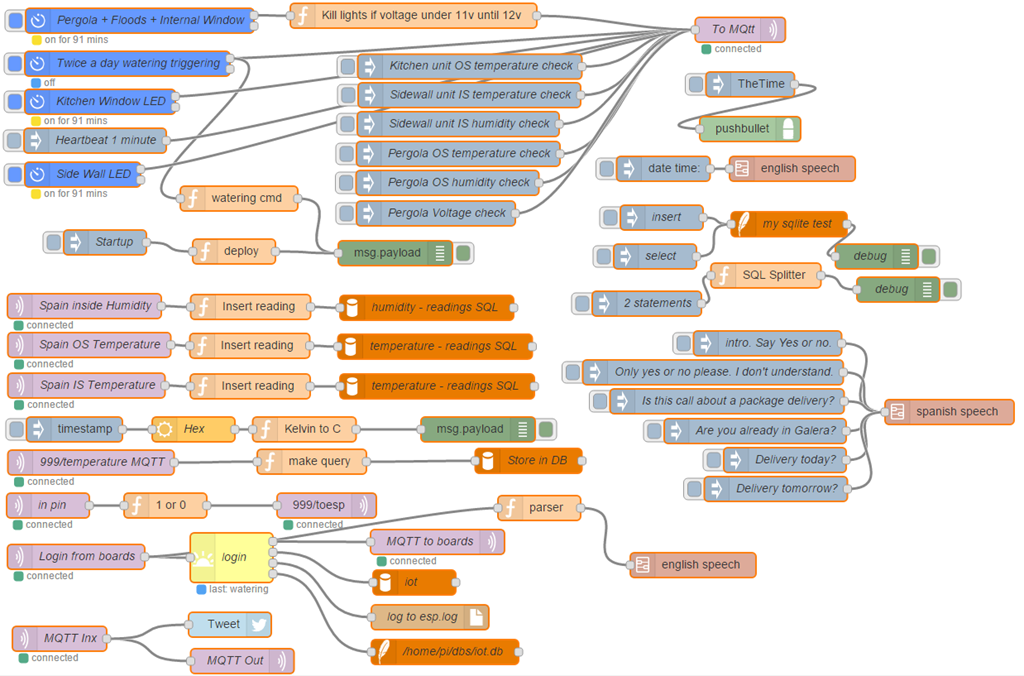}
	\caption[A complex Node-RED \textit{flow}]{An example of a complex Node-RED \textit{flow}~\cite{scargill_2015}.}
	\label{fig:noderedmad}
\end{figure}

However Node-RED development tool has some limitations such as not having a proper mechanism in order to debug and/or test the developed \textit{flows}. Given that IoT systems are typical of large-scale and complex by nature, it is easy to end-up with \textit{flows} as complex as the one given on Figure~\ref{fig:noderedmad}, thus we can observe that it does not scale, at least from a developing process perspective.

It is also noticeable that it does not leverage the use of models as way of abstracting components (e.g. devices), thus increasing the complexity of developing new \textit{nodes} both on dealing with the essential complexity of the \textit{node} (e.g. a new algorithm) and in the accidental complexity of communicating with the already existent nodes.

Further, there is no liveness in the development of IoT solutions with Node-RED due to the need of \textit{deploy} each time as the new modification is made on a given \textit{flow}, reducing the feedback-loop to the developer (that must use \textit{logging} mechanisms in order to observe the correct functioning of the system).

\subsection{Visual Programming for IoT}

Diagrams and other graphical logic and/or model representations have been playing a role in software development since the appearance of the modern digital computers in the 1940s. In the beginning, the diagrams were paper-based aids, used to design and understand the software structure, but then, interest appeared in the direct use of diagrams as a solution to improve software development tools. This led to the appearing of visual software project management tools, visual editors for graphical interface creation, visual tools for software modeling and engineering, and visual programming languages~\cite{vplbook}.

A \textit{Visual Programming Language} (VPL) can be defined, as described in the \textit{Wiley Encyclopedia of Computer Science and Engineering}~\cite{S.K.Chang202}, as:
\begin{displayquote}
\textit{A language in which significant parts of the structure of a program are represented in a pictorial notation, which may include icons, connecting lines indicating relationships, motion, color, texture, shading, or any other non-textual device.}
\end{displayquote}

As such, visual programming makes use of an extensive set of icons and diagrams to convey information and to allow multi-modal communication and interaction between humans and computers~\cite{S.K.Chang202}. 

VPLs have been explored and used in several domains, including, but not limited to, educational activities (e.g. learning to program), multimedia, video game development, system design and development, simulations, automation, data warehousing and business analytics~\cite{Ray2017}.

Although several domains of applications already take advantage of the use of VPLs, the emerging field of Internet-of-Things (IoT) is still lingering far behind other sectors.

Visual Programming Languages are commonly used with the intent of abstracting low-level concepts and details into a more high-level logic, through the use of visual metaphors ~\cite{Thomas:2003:MDD:949344.949346,Barricelli2015}. The application of domain-specific visual programming languages to solve the need of abstraction from the low-level and heterogeneous devices that usually make a part in the IoT connected world can already be observed. We can consider, as an example, how graphical-based programming languages are already widely used for programming low-level devices such as Programmed Logic Controllers (PLCs) in high-production manufacturing systems~\cite{Younis2003}.

The goal of this section is to review the landscape of visual programming environments (\textit{viz.} Integrated Development Environments or other kind of development toolkits) and their inter-winded visual programming languages (which can be novel or only an adaptation of an already existent visual language) in the context of the IoT and characterize them. In order to do so, a novel characterization methodology was developed based on previous work on characterizing VPLs, with considerations about the IoT domain and the development toolkits associated with the VPLs.

The paper is structured as follows: in Section \ref{categories} it is given an overview of the literature about characterizing VPLs. Section \ref{proposal} presents a proposal on how to characterize Visual Programming for IoT (languages and supporting development toolkits). An overview of the available languages for IoT is given on Section \ref{catalog} and a comparison of those is given on Section \ref{comparison}. Some final remarks are given on Section \ref{concl}.

\subsubsection{Characterizing VPLs} \label{categories}
\label{S:2}

VPLs are problematic to characterize and classify due to the variety of formalisms used to define them. However, there are some key concepts shared in the available literature. As an alternative to traditional programming, VPLs try to primary improve four aspects of programming~\cite{Johnston2004,Burnett1994287}:
\begin{itemize}
\item \textit{Simplicity}: Increase the simplicity of programming tasks by the means of reducing the key concepts needed to construct and understand programs;
\item \textit{Concreteness}: Allow the direct and visual exploration of data;
\item \textit{Explicitness}: Explicit definition of relationships between the elements of the program;
\item \textit{Responsiveness}: The changes to the program are immediately visible, giving immediate feedback.
\end{itemize}

As of today, there is a wide range of VPLs available, meeting different uses and system requirements. As so, these languages typically present different characteristics, paradigms, and features, allocating themselves in one or more classes~\cite{Burnett1994287}. We can classify VPLs in the same terms as traditional programming paradigms, for example, as imperative, declarative, functional, logic or object-oriented. 

\subsubsection{Visual Metaphors}

Brunett et al.~\cite{Burnett1994287} were one of the first to define a list of eleven main paradigms present in VPL that include the traditional paradigms in programming but also others. Based on their classification, we can consider that there are two main approaches for developing VPLs, namely: (1) \textit{graph-based}, which are the most disseminated and in some way leverage the metaphor of visual graphs (nodes and connections), and, (2) \textit{box-based}, which leverage the use of the metaphor of box and sub-boxes (e.g. User-Interfaces builders and Forms). 

\subsubsection{Paradigms}

We can consider the next set of paradigms as the main paradigms as they are the most widespread in the literature~\cite{book:55167,S.K.Chang202,Burnett1994287}:

\begin{itemize}
\item \textbf{Data-flow languages}: Computation is specified by the means of graphs. This graph consists of icons (or similar visual representations) that correspond to operational nodes, being these nodes connected by lines that represent the flow of data between them.
\item \textbf{Component-based languages} (\textit{graph-based}): These languages are based on the metaphor of networked computing devices or \textit{components}. Each one of those \textit{components} can perform a variety of tasks in response to messages and data received from others.
\item \textbf{Rule-based languages}: Language based on the definition of triggering actions that happen upon changes (e.g. the modification of a variable value can trigger an action). Usually, these mechanisms are based upon visual \textit{if-then} rules.
\item \textbf{Program-by-demonstration languages}: Languages that, instead of relying on the specification of actions through instructions or commands, depend on demonstrations of the pretended task or objective. As so, the language bases itself on the manipulation of visual objects, and the tasks carried out are performed according to those manipulations.
\item \textbf{Form-based \& Spreadsheet-based languages}: Spreadsheets are the most widespread VPL paradigm. These languages present a ledger-like sheet for entering and performing arithmetic on values. In these languages, the sheet is the single significant pictorial element that qualifies the original spreadsheet as a VPL. Form-based languages are somewhat similar in the scene that they result in a generalization of sheets into forms.
\end{itemize}

\subsubsection{Features}

In spite of the variety of paradigms, it must be noticed that these are not mutually exclusive since one single language can fit in more than one. In addition to these, VPLs take advantage of a set of different features as an extension to them, as a way of improving the language capabilities and the developer's experience. With this in mind, we can consider the following as the most relevant features commonly found in VPLs~\cite{Boshernitsan2004,Burnett1994287}.

\subsubsection*{Abstraction} The two most widely supported types of abstraction are \textit{procedural abstraction} and \textit{data abstraction}. Regarding \textit{procedural abstraction}, it can be split into two levels: high-level and low-level. On one hand, high-level visual programming languages, found in various domain-specific systems such as software maintenance tools and scientific visualization environments, are not complete programming languages in such way that it is not possible to write and maintain an entire program on it, always depending on some sort of underlying non-visual modules. On the other hand, low-level VPL does not allow the programmer to combine fine-grained logic into procedural modules. Typically, VPLs (especially general-purpose ones) combine low-level and high-level abstractions. \textit{Data abstraction} facilities are only found in general-purpose programming languages, and, for this kind of abstraction, the requirements are that data types are defined visually, have a visual representation and provide interactive behavior~\cite{Boshernitsan2004}.

\subsubsection*{Control-flow} As in conventional programming languages, VPLs embrace two notions of control-flow: (1) imperative or (2) declarative. For the (1) imperative approach, there are one or more flow diagrams which indicate how the thread of control flows through the program, with the advantage of, for example, having an effective visual representation of parallelism. In this case, the developer is required to keep track of how sequencing of operations modifies the program state. In its counterpart, (2) declarative approach, there is only the need to worry about what computations are performed, and not how the actual operations are carried out. In this case, explicit state modification is avoided by using single assignment: instead of modifying an existent object's state, the programmer copies an existent one and, then, specifies the desired differences~\cite{Boshernitsan2004}.

\subsubsection*{Event and Exception Handling} Event handling deals with events triggered by changes in an object state, hardware interrupts or user interaction. Possible implementations of such mechanism are single event structure (one event at a time) or multiple event structures (event switch technique and/ or dynamic computation node). Exception handling follows the same principle but responds to exceptions in the system. Both can be implemented in form of icons or any other visual way~\cite{Boshernitsan2004}.

\subsubsection*{Visual Structures and Data Types} Most VPLs rely on visual representations of structures and data types. These representations follow the same basic concepts behind primitive data types, but with different visual representations. Additionally, the programmer is, in some cases, allowed to create new data types by the means of inheritance and/or encapsulation~\cite{Boshernitsan2004}.

\subsubsection{Categories}

More recently, and based on the different \textit{paradigms} and \textit{features} that VPLs have, Boshernitsan et al. defined five major and broader categories for classifying VPLs~\cite{Boshernitsan2004}, based on the previous work of Chang, Shu, and Burnett~\cite{book:55167,S.K.Chang202,Burnett1994287}. As such, nowadays, the five categories more used to characterize VPLs are the following:
\begin{itemize}
\item \textbf{Purely Visual Languages}: Languages that totally rely on visual techniques throughout the programming process and the program is compiled directly from its visual representation. In this case, the language is never translated into an interim text-based language.
\item \textbf{Hybrid Systems (Textual/Visual)}: Languages that can be either created visually and then translated into an underlying high-level textual language or involve the use of graphical elements in an otherwise textual language.
\item \textbf{Programming-by-Example systems}: In these systems, the paradigm of programming-by-demonstration is followed and the user is allowed to create and manipulate visual objects in order to \textit{teach} the system how to perform tasks.
\item \textbf{Constraint-oriented Systems}: Languages that are designed to act in a constraint scenario or environments, such as simulation design or graphical user interfaces development.
\item \textbf{Form-based systems}: Languages that take advantage of any spreadsheet metaphor.
\end{itemize}

It is important to note that these categories are by no means mutually exclusive. Indeed, many languages can be placed in more than one category.

\subsubsection*{Characterizing IoT VPLs and their Development Environments} \label{proposal}
\label{S:3}

A typical development lifecycle for the IoT is similar to the development of any other system, plus some particularities inherent to the IoT ecosystem. However, since IoT is still in its early stages of development, its development mechanisms and tools are still lagging behind on the best practices and lessons learned from the Software Engineering community in the past decades. Such earliness of development is noticed in the lack of Integrated Development Environments with proper mechanisms of debugging and testing~\cite{VVioT2018}. 

In spite of the novelty of the IoT, several Visual Programming Languages appeared (altogether with their supporting tools and development environments). And, similarly to any other programming language, there is an array of aspects that have impact when it comes to picking the Visual Programming system to use, beyond the language itself, but the features it provides and others aspects such as the openness, extensibility, community support, thus, we see that the traditional characterization approaches do not cover all the details of these systems.

In the following paragraphs we propose a set of guidelines for characterizing these languages and their development environments, enumerating the aspects that must be taken into account. This proposal bases itself on several core aspects that were considered as unique between the different VPL solutions available, as follows:

\begin{table}[h]
\centering
\resizebox{\linewidth}{!}{%
\begin{tabular}{l l l l l l l l}
\hline
\textbf{IoT Layer} & \textbf{\begin{tabular}[c]{@{}l@{}}Abstraction \\ Level\end{tabular}} & \textbf{Category} & \textbf{Control-Flow} & \textbf{License} & \textbf{\begin{tabular}[c]{@{}l@{}}Target\\ Platforms\end{tabular}} & \textbf{Extensible} & \textbf{\begin{tabular}[c]{@{}l@{}}3rd-party \\ Integration\end{tabular}} \\ \hline
\begin{tabular}[c]{@{}l@{}} Edge\\ Fog\\ Cloud\end{tabular} & \begin{tabular}[c]{@{}l@{}} Data\\ Procedural\end{tabular} & \begin{tabular}[c]{@{}l@{}} Hybrid\\  Constraint-oriented\\ Purely Visual\\ Form-based \\ PBE\end{tabular} & \begin{tabular}[c]{@{}l@{}} Imperative\\  Declarative\end{tabular} & \begin{tabular}[c]{@{}l@{}} Open-source\\  Close-source\end{tabular} & \begin{tabular}[c]{@{}l@{}} Linux-based\\  Arduino-based\\  (others)\end{tabular} & \begin{tabular}[c]{@{}l@{}} Yes (Extensible)\\  No (Not extensible)\end{tabular} & \begin{tabular}[c]{@{}l@{}} Full\\ Partial\\  None\end{tabular} \\ \hline
\end{tabular}%
}
\caption{Summary of the characterization scheme for IoT VPLs.}
\label{table:class}
\end{table}

\begin{itemize}
\item \textbf{IoT Layer}: Each VPL can target one or more layers of the IoT system design. As such, VPLs can be categorized by its target layer:
\begin{itemize}
\item Edge, Fog, Cloud.
\end{itemize}
\item \textbf{Abstraction Level}: In accordance with the state-of-the-art on characterizing VPLs (Section \ref{categories}), VPL can be distinguished based on the levels of abstraction, fitting in one of the following:
\begin{itemize}
\item Procedural abstraction, Data abstraction. 
\end{itemize}
\item \textbf{Category}: In accordance with the work of Boshernitsan et al.  (Section \ref{categories}), VPL can fit in some broader categories depending on the features they provide and the inherent paradigm they use, fitting in one of the following:
\begin{itemize}
\item Purely-Visual, Hybrid, Constraint(-oriented), Form(-based), PBE (Programming-by-Example)
\end{itemize}
Sub-categories are given based on the programming paradigm that the languages follow (Section \ref{categories}).
\item \textbf{Control-Flow}: In accordance with the state-of-the-art on characterizing VPL (Section \ref{categories}), VPL can be distinguished based on the control-flow they use, fitting in one of the following:
\begin{itemize}
\item Imperative, Declarative.
\end{itemize}
\item \textbf{Target Platforms}: VPLs can target the development of different platforms and/or operative systems. Thus, each VPL can target one or more of the following systems:
\begin{itemize}
\item Linux(-based systems), Arduino(-based systems), Others (e.g. Java-based, Android).
\end{itemize}
\item \textbf{License}: VPLs can be proprietary software or open-source, being this an aspect of increasing importance highly due to the high-heterogeneity of the IoT systems and devices and the constant needs of making case-by-case modifications on the VPL itself. As such, a VPL can be:
\begin{itemize}
\item Open(-source), Close(-source).
\end{itemize}
\item \textbf{Extensibility}: VPL's can allow the development of extensions, for example, add-ons or add support to new/different hardware or software stacks, a feature that can increase or reduce the applicability of the language in the high-heterogeneous IoT ecosystem. So, the VPLs can be:
\begin{itemize}
\item Extensible, Not extensible.
\end{itemize}
\item \textbf{3\textsuperscript{rd}-party Support}: VPLs can support 3\textsuperscript{rd}-party integration out-of-the-box, allowing interaction with other systems by the means, for example, of application programming interfaces (API's). This support can be:
\begin{itemize}
\item Full, Partial, None.
\end{itemize}
\end{itemize}

This proposal of characterization of VPLs in IoT can be seen as a foundation for classifying such languages, identifying open research directions, issues or lacks within the available languages as today. It is of importance to note that each characterizing variable can have more than one value. A summary of the classification scheme can be seen on Table \ref{table:class}.

\subsubsection{IoT Visual Programming} \label{catalog}

There are several solutions for developing IoT-based systems that leverage Visual Programming Languages. These solutions are mostly distinct, having a different focus, set of feature and base themselves in different paradigms. An overview comparison and discussion of the tools available is given in Section \ref{comparison}. 

The tools have been selected after a curated search on scholar database (Scopus) and Google search engine, and are shortly described in the following paragraphs. The keywords used for the search were one or a combination of the following: \texttt{Internet-of-Things}, \texttt{VPL}, \texttt{Visual Programming}, \texttt{Visual Programming Language}, \texttt{Visual} and \texttt{IoT}.

\begin{figure}[h]
    \centering
    \includegraphics[width=0.95\textwidth]{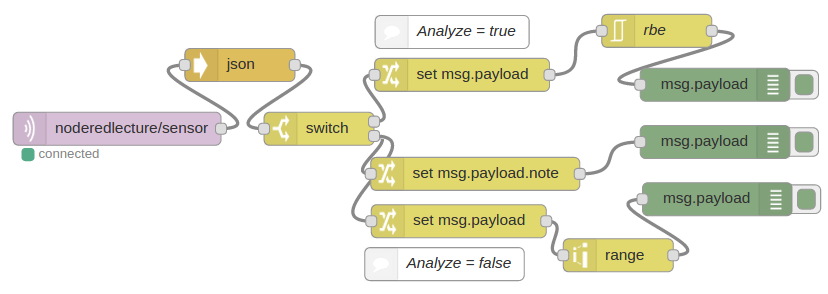}
    \caption{An example of a Node-RED Flow.}
    \label{fig:nodered}
\end{figure}

\begin{description}

\item [Node-RED]~\footnote{Website: https://nodered.org/}

Node-RED is a IoT-focused development toolkit that presents a flow-based development for wiring together hardware devices, APIs, and online services. It was originally developed by IBM, and its runtime is built on Node.js\footnote{Node.js: https://nodejs.org}. The language can be extended by building new blocks using JavaScript snippets. The flows created can be exported in JSON format. A sample flow diagram is given in Figure~\ref{fig:nodered}.

\item [Flogo]~\footnote{Website: http://www.flogo.io/}

Project Flogo is an Open Source Framework for IoT Edge Apps and Integration. It leverages the use of \textit{flows} (available thought its Flogo Web UI) as a visual programming paradigm for development.

\item [NETLab Toolkit]~\footnote{Website: https://www.netlabtoolkit.org/}

NETLab Toolkit (NTK) is a visual programming environment that empowers designers, developers, makers, researchers, and students who want to design and build tangible IoT projects. Its visual language allows the connection of sensors, actuators, media and networks with \textit{drag-and-drop} smart widgets.

NTK works with Arduino and Linux-based embedded systems (e.g. Intel Edison, Raspberry Pi). The language can be easily adapted for new \textit{things}, by allowing the user to develop their own widgets using Javascript snippets.

\begin{figure}[h]
    \centering
    \includegraphics[width=0.75\textwidth]{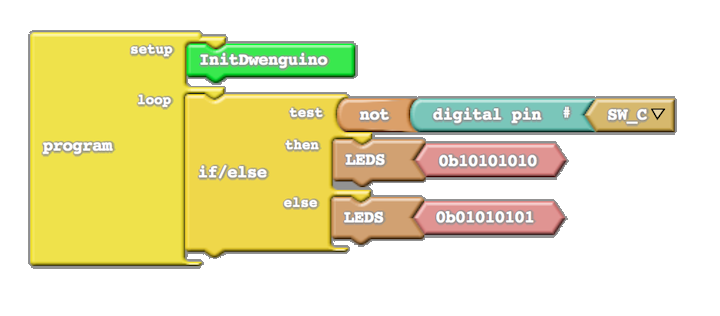}
    \caption{An example of a ArduBlock code snippet.}
    \label{fig:ardublock}
\end{figure}

\item [ArduBlock]~\footnote{Website: https://ardublock.com/}

ArduBlock is a programming environment designed to allow \textit{drag-and-drop} programming of physical computing devices that use Arduino. The visual language allows the use of visual code-block and connections between them as a way of programming. An example of a code-snippet is given in Figure \ref{fig:ardublock}.

\item [S4A]~\footnote{Website: http://s4a.cat/}

S4A is a modified version of the original Scratch programming language (focused on educational purposes) that targets Arduino-based hardware. It provides visual blocks as a way to manage sensors and actuators connected to Arduino-based boards. It keeps the Scratch original focus on educational purposes only.

\item [Modkit]~\footnote{Website: http://www.modkit.com/}

Modkit for VEX is a graphical programming environment developed specifically for VEX IQ~\footnote{VEX IQ is a robotics platform used for teaching the basics of programming and robotics.}.
It is based on the Scratch programming language, basing itself on snap-together visual blocks, and maintain the focus on educational purposes.

\item [miniBloq]~\footnote{Website: http://minibloq.org}

miniBloq is a visual programming environment targeting Multiplo boards, Arduino boards, physical computing devices, and robots. miniBloq presents itself as an all-in-one language for interaction with hardware. It is highly modular, allowing the developer to create new and personalized blocks. Plus it allows the developer to easily add support to new boards and platforms by the extension of the given language.

\item [NooDL]~\footnote{Website: https://www.getnoodl.com/}

NooDL is a visual development environment that allows designers and developers to visually create interfaces, logic and data flows. Although it does not focus on IoT, it also covers IoT based system programming, and it is based on nodes, connections, and hierarchies in order to do so.

\item [DGLux5]~\footnote{Website: http://www.dglogik.com/}

DGLux5 is a \textit{drag-and-drop} rapid application development and visualization platform, that allows the development of real-time, data-driven applications and dashboards. The language used allows the unification of different data systems and multiple data providers into a single interface. It provides sample workflows.  

\item [AT\&T Flow Designer]~\footnote{Website: https://flow.att.com/}

AT\&T IoT Platform Flow IDE is a cloud IoT focused development environment. The visual language allows the creation of prototypes of IoT solutions, giving the ability to iterate and improve through multiple versions, then deploy the final solution. It gives the developer a set of preconfigured nodes that allow easy access to multiple data sources, cloud services, device profiles, and communication methods. It is based upon the Node-RED programming environment.

\item [Reactive Blocks]~\footnote{Website: http://www.bitreactive.com/}

Reactive Blocks is a visual model-driven development environment supporting formal model analysis, automated code generation, hierarchical modeling, and an extensive library of ready-to-use components for the Java platform. By combining re-usable blocks, a developer can create complex applications graphically. Although the visual language does not focus on IoT development, it allows the integration with any Java-based IoT stack and  OSGi based IoT stacks, allowing the dynamic starting, stopping, and re-configuration of applications.
 
\item [GraspIO]~\footnote{Website: https://www.grasp.io/}

GraspIO Graphical Smart Program for Inputs and Outputs is a part hardware and part software platform that offers the user the ability to quickly build IoT and Robotics systems. However, the tool is yet under development, so there is lack of details about the visual language used.

\item [Wyliodrin]~\footnote{Website: https://www.wyliodrin.com/}

Wyliodrin is an online IDE for Linux-based embedded systems (Raspberry Pi, Intel Galileo) programming. It provides a \textit{drag-and-drop} visual programming language to interact with the hardware, and it is easily extensible using Python or JavaScript.

\item [Zenodys]~\footnote{Website: https://www.zenodys.com/}

Zenodys is a fully visual IoT platform for Industry 4.0. It provides a visual development console, ZenoVisual, that allows for visual program workflows to deal with the data coming from IoT devices and integrate with third-party applications or APIs. The developed system can run in the cloud or on-premises.

\item [ASU VIPLE]~\footnote{Website: https://neptune.fulton.ad.asu.edu/VIPLE/} \cite{chen2016viple}

Visual IoT/Robotics Programming Language Environment (VIPLE) bases itself on Microsoft Robotics Developer Studio\footnote{Microsoft Robotics Developer Studio support and development has been discontinued by Microsoft.} and extends its functionalities. It leverages the use of a visual programming language to develop a variety of IoT systems and robotics platforms, with an open programming API and interface.

The VIPLE program runs on a backend PC, and receives sensor and motor feedback, and sends commands to the robot motors, and supports both Bluetooth and Wi-Fi connections (communicating using JSON objects).

\begin{figure}[h]
    \centering
    \includegraphics[width=0.75\textwidth]{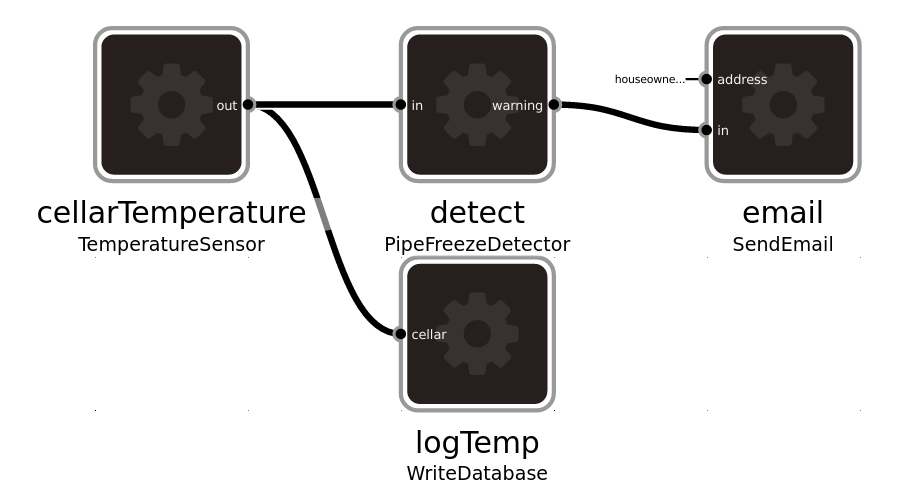}
    \caption{An example of a data-flow in the Flowhub IoT Platform.}
    \label{fig:flowhub}
\end{figure}

\item [Flowhub IoT Platform]~\footnote{Website: https://flowhub.io/iot/}

Flowhub is a web-based IDE for flow-based programming. The visual language allows the use of visual blocks and connections as a way of programming the system's logic. It is built on NoFlo.js for both client and server and can connect to 3rd party systems using the FBP Network Protocol. An example of a data-flow in the platform is given in Figure \ref{fig:flowhub}.

\item [XOD]~\footnote{Website: https://xod.io/}

XOD is an edge device programming platform (microcontrollers). It uses a visual language to programming the devices and then generates native code for the target platform. In the language, a node is a block that represents either some physical device like a sensor, motor, or relay, or some operation such as addition, comparison, or text concatenation. Each node has one or more inputs that accept values to be processed and outputs that return results. Creating a link from an output to an input builds a path for data, allowing one node to feed values into another. 

\begin{figure}[ht]
    \centering
    \includegraphics[width=0.75\textwidth]{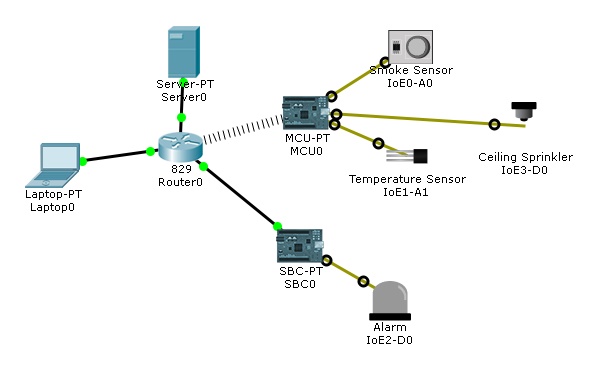}
    \caption{An example of a Packet Tracer 7.0 visual diagram.}
    \label{fig:pt7}
\end{figure}

\item [Packet Tracer 7.0]~\footnote{Website: http://www.packettracernetwork.com/} 

Cisco Packet Tracer is a tool for network simulations that use a visual paradigm as a way of identifying the different devices and connections. Although it does not allow the development of IoT solutions, it provides a visual language that allows the prototyping of IoT architectures and systems. An example diagram is given in Figure \ref{fig:pt7}.

\item [Visuino]~\footnote{Website: https://www.visuino.com/}

Visuino is a  visual programming environment for Arduino-based boards. It provides a \textit{drag-and-drop} visual programming language to interact with the hardware, and then make the connection between the different modules.

\item [WoTKit]~\footnote{Website: https://sensetecnic.com/}~\cite{blackstock2012iot}

WoTKit provides a cloud-hosted Node-RED service (FRED) to allow the development of \textit{cloud layer} IoT solutions. It follows the same visual programming principles of Node-RED and additional extensions and integrations. The solutions are part of the STS IoT Platform that is tightly integrated with FRED.

\item [Losant]~\footnote{Website: https://www.losant.com}

Losant is a developer platform that provides scalable device management, data collection, data visualization, and data reaction workflows in real-time. As such, the implemented visual language is based on workflows, allows the definition of how the devices communicate with each other and other services, focusing on the data component of the IoT.

\item [IFTTT]~\footnote{Website: https://ifttt.com/}

IFTTT is a mobile application that leverages the use of a visual programming language in order to develop \textit{if-this-then-that} rules. The language provides integration with several third parties. Despite not being focused on IoT solutions, it provides integration with several IoT products on the market allowing the programming of their behavior.

\item [Blynk]~\footnote{Website: https://www.blynk.cc/}

Blynk is a mobile application that allows the control of Arduino and Raspberry Pi devices using a digital dashboard. It allows the \textit{drag-and-drop} of widgets and then configures them using a form-like approach.

\item[EPIDOSITE]~\cite{10.1007/978-3-319-58735-6_1}

EPIDOSITE is a mobile programming-by-demonstration system focused on End-User Development (EUD). It extends the prior mobile PBD system Sugilite~\cite{li2017sugilite}, using the Android’s accessibility API to support automating tasks in Android apps, by recording all of the user's interactions with the phone, together with the relevant UI elements on the screen while using different IoT applications, and, then processes the recording and generates a reusable script for performing the task. EPIDOSITE extends Sugilite by adding new features and mechanisms to support the programming of IoT devices (e.g. new ways for triggering scripts, external services and devices).

\end{description}

\begin{table}[ht]
	\centering
	\begin{adjustbox}{width = \textwidth,max totalheight=\textheight}
		\begin{tabular}{P{1.8cm} P{1cm} P{2cm} P{2cm}  l l P{2.2cm} P{1.3cm} P{1.95cm}}
			\toprule
			VPL & IoT Layer  & Abstraction Level        & Category                    & Control-Flow & License & Target Platforms            & Extensible & 3rd Party Integration \\ \midrule
			\textbf{Node-RED}             & Any        & Data, Procedural         & Hybrid                      & Declarative  & Open        & Linux-based                 & Yes        & Full                   \\ \hline
			\textbf{Flogo}                & Fog        & Data, Procedural         & Hybrid                      & Declarative  & Open        & Linux-based                 & Yes        & Full                   \\ \hline
			\textbf{NETLab Toolkit}  & Edge       & Procedural         & Hybrid                     & Imperative   & Open        & Arduino-based, Linux-based  & Yes        & Yes                   \\ \hline 
			\textbf{Ardublock}            & Edge       & Procedural         & Constraint-oriented         & Imperative   & Open        & Arduino-based               & No         & None                    \\ \hline
			\textbf{S4A}                  & Edge       & Procedural         & Constraint-oriented         & Imperative   & Open        & Arduino                     & No         & N/A                   \\ \hline
			\textbf{Modkit}               & Edge       & Procedural         & Constraint-oriented         & Imperative   & Close      & Arduino, VEX                & No         & None                    \\ \hline
			\textbf{miniBloq}             & Edge       & Procedural  & Hybrid, Constraint-oriented & Imperative   & Open        & Arduino, Multiplo, Others        & Yes        & N/A                   \\ \hline
			\textbf{NooDL}                & Cloud      & Data             & Purely Visual                 & Declarative  & Close      & N/A                         & No         & Full                   \\ \hline
			\textbf{DGLux5}               & Fog, Cloud & Data            & Purely Visual                & Declarative  & Close      & N/A                         & Yes        & Full                   \\ \hline
			\textbf{AT\&T Flow}  & Cloud      & Data, Procedural     & Purely Visual                & Declarative  & Close      & N/A                         & No         & Full                   \\ \hline
			\textbf{Reactive Blocks}      & Fog, Cloud & Procedural           & Hybrid                       & Declarative  & Close      & Java-based                  & Yes        & Full                   \\ \hline
			\textbf{GraspIO}              & Edge       & Procedural            & Constraint-oriented          & Declarative  & Close      & Linux-based, Arduino        & N/A        & N/A                   \\ \hline
			\textbf{Wyliodrin}            & Any        & Data, Procedural   & Constraint-oriented         & Declarative  & Close      & Linux-based, Arduino        & Yes        & Full                   \\ \hline
			\textbf{Zenodys}              & Cloud      & Data, Procedural              & N/A                         & Declarative  & Close      & N/A                         & Yes        & Yes                   \\ \hline
			\textbf{ASU VIPLE}            & Edge       & Procedural               & Hybrid                        & Imperative   & Open        & EV3, Intel-based, ARM-based & No         & None                    \\ \hline
			\textbf{Flowhub IoT} & Fog, Cloud & Data, Procedural       & Hybrid                    & Declarative  & Open        & Linux-based                 & Yes        & Yes                   \\ \hline
			\textbf{XOD}                  & Edge       & Procedural               & Purely Visual                 & Declarative  & Open        & Arduino-based               & Yes        & N/A                   \\ \hline
			\textbf{Packet Tracker}       & Fog        & N/A                & Constraint-oriented           & N/A          & Close      & N/A                         & Yes        & N/A                   \\ \hline
			\textbf{Visuino}              & Edge       & Procedural              & Purely Visual                  & Imperative   & Close      & Arduino                     & Yes        & None                    \\ \hline
			\textbf{WoTKit}               & Cloud      & Data, Procedural        & Hybrid                         & Declarative  & Close      & N/A                         & Yes        & None                   \\ \hline
			\textbf{Losant}               & Cloud      & Data                   & Purely Visual                 & Declarative  & Close      & N/A                         & Yes        & Full                   \\ \hline
			\textbf{IFTTT}                & Cloud      & Data                   & Purely Visual               & Declarative  & Close      & Android                     & No         & Partial                   \\ \hline
			\textbf{Blynk}                & Any        & Data                   & Form-based                  & Declarative  & Open        & Android, iOS                & No         & Full                   \\ \hline
			EPIDOSITE    & Any        & Procedural                   & PBE                  & Declarative  & Open        & Android                & N/A       & N/A                   \\
\bottomrule
		\end{tabular}
	\end{adjustbox}
	\caption[IoT Visual Programming Environments comparative overview]{Internet-of-Things Visual Programming Environments and their coupled Visual Programming Languages comparative overview\footnote{N/A stands for information not available or not applicable.}. The characterization is based on the proposal of Section \ref{categories}.}
	\label{table:vpls}
\end{table}

\subsubsection{IoT VPLs Comparison} \label{comparison}

As of today, there are already a considerable number of VPLs on the market, within the scope of IoT programming. A comparative overview of these languages is given on Table~\ref{table:vpls}. For most of the languages, the name presented in the Table~\ref{table:vpls} belongs to the IDE or platform that supports the language and not to the language itself.

Firstly, and based upon the division of languages per IoT layer, it can be observed that there is a balance in the number of solutions available for each one of them. However, most of the solutions targeting the \textit{fog layer} can also be used in the \textit{cloud layer} (since the majority of them targets Linux-based systems, common to both the \textit{fog layer} and the \textit{cloud layer}). Additionally, it is noticeable that the languages that target the \textit{edge layer} (devices) are very specific to this layer only (mostly due to the direct interaction with hardware inputs/outputs).

Regarding the abstraction level, most of the languages have \textit{procedural abstraction}, both at a high-level and low-level. Although there's a small number that focuses on \textit{data abstraction}, and for the ones that do, they typically focus on higher layers of the IoT architecture.

Most of the languages are based on a \textit{data-flow} or \textit{component-based} paradigm. Given that most of the IoT systems deals with, on one hand, data collections and transformations, and, on the other hand, the interaction between different components' abstractions, from hardware to software based ones. The languages more directed to user-level interaction, like the \texttt{IFTTT}, follow the \textit{rule-based} paradigm, since it is of easy understanding.

Regarding the categories on which each VPL fits, there is a good distribution, from \textit{hybrid} ones such as the most known \texttt{Node-RED}, \textit{constraint-based} ones, generally targeting the \textit{edge layer}, \textit{purely visual} that are common on languages used on the \textit{cloud layer}. With less expression the \textit{form-based} ones appear only to be used by the \texttt{Blynk} platform.

At the \textit{control-flow} perspective most of the languages that target the \textit{edge layer} have an \textit{imperative} \textit{control-flow}, similarly to the non-visual languages used to this layer, which are also imperative (e.g. Arduino language). More higher-level languages, such as the \texttt{Node-RED}, follow a \textit{declarative} philosophy.

From a developer-focused perspective, the openness of the solutions highly impacts the development process, allowing the improvement/addition of features to the internals of the language. In the context of VPLs for IoT, there's a balance between open-source and closed source solutions. Although, there are some close-source languages that are based upon open-source solutions such as, e.g., the Scratch language\footnote{Scratch Website: https://scratch.mit.edu/}.

At the target-platform for each one of the languages, it is noticeable that languages that target the \textit{edge layer} are more limited and depend on specific hardware. However, most of the other languages target Linux-based operating systems which are highly widespread in the market for Fog devices (e.g. Raspberry Pi) and cloud systems.

Other useful aspect is the ability that a language has in terms of extensibility and 3rd party integration support. It is noticeable that most of the languages have both the support of extensions of some sort and are easily integrated with 3rd party systems.

In summary, the IoT VPLs characteristics are well distributed, with a good amount of different solutions targeting different scopes of application. Although there are open-issues, namely regarding on how to deal with the heterogeneity, mainly in the \textit{edge layer}, and on how these languages scale with the increase of complexity of the IoT systems.

\subsection{Summary}

As Prehofer et al. points on their work \textsc{From Internet of things mashups to model-based development}~\cite{Prehofer2015} (and aforementioned in Section~\ref{ssec:architecturestyles}), there are two kinds of tools for developing IoT systems, namely: mashup approaches and model-based approaches. Hitherto were described two of the most known tools that fall into these two types, respectively, Node-RED and ThingML.

From the analysis of this two tools, several conclusions were made, showing lacks on both. Node-RED as a mashup tool has several lacks in what regards the leverage of using models (and models@runtime) or anything related to \textit{liveness}. However, it has a simple to use drag-n-drop visual programming interface that can be used as a reference in the domain of IoT, but needs to be extended to embrace the necessities of \textit{liveness} and its feedback-loop needs.

ThingML does a full leverage of models as a way to developed IoT systems, however, it does not embrace the use of visual notations as a way of developing these systems, being text-based by default. Further, as aforementioned, it has just a preliminary study on how to embrace models@runtime, so it was not designed from the ground up to embrace this concept which leads to several limitations.

In order to have a development environment tailored for the IoT that leverages models and live to programme by design, the lessons learned and features that are proven to work (such as the visual notation of Node-RED) from these systems can and must be leveraged in a new solution. However, since all the components on these systems are highly inter-winded, it is not viable to use any of this solutions as the foundation for a new development solution.


This paper focuses on giving an overview of the available solutions in the landscape of visual programming languages for the Internet-of-Things. An introduction to the field of the IoT and visual programming is given, and the applicability of such programming approach in the IoT domain.

An overview of the literature on characterizing VPLs is provided, but there are gaps in literature when approaching the specificities of the IoT scenario. As such, a proposal for characterizing IoT VPLs is given, based on the existent literature, plus some aspects that are of importance in IoT development.

A compendium of the available visual programming languages targeting IoT development was elaborated and an overall comparison of the available solutions is provided, using the proposed classification guidelines. 

With this comparison it was observed that there is a good amount of solutions available, covering different aspects of the development. However, some limitations were found, especially on the \textit{edge} development and scalability perspective. 

There is still some research challenges that must be addressed by the community on the improvement of the existent languages or on the creation of new ones in order to reach the full potential of Visual Programming for the Internet-of-Things.


\section{Testing the Internet-of-Things}

The Internet-of-Things relies on a combination of hardware and software that enable real-world objects to sense and interact with the surrounding environment while being Internet-connected and uniquely identifiable~\cite{whitmore2015internet}. As such, in order to guarantee IoT-based system's performance, scalability, reliability, and, further, security, it is needed focus on testing the different layers and components that make part of the system, from low-level/hardware specifications to high-level components. It is hard to draw a line between the low-level and high-level components in IoT-scope since they are strongly connected and dependent, however, the methods and techniques used for testing these are, typically, similar.

\subsection{Testing Levels and Methods}

Testing approaches can be in one or more levels, depending on the scope of the test and objective. So, different test levels are defined, as follows~\cite{beizer2003software}:
\begin{description}[align=left]
	\item[Unit Testing] Testing of individual hardware or software units or groups of related units~\cite{159342}. It consists of isolating each part of the system and shows that individual parts fit its requirements and functionalities.
	\item[Integration Testing] Software and/or hardware components are combined and tested to check the interaction between them and how they perform together~\cite{159342}.
	\item[System Testing] Testing a complete, integrated system to check the system's compliance and behaviour within the specified requirements~\cite{159342}
	\item[Acceptance Testing] Formal testing conducted to determine whether or not a system satisfies its acceptance criteria and to enable a customer, a user, or other authorized entity to determine whether or not to accept the system~\cite{159342}
\end{description}

Different methods can be used to test the \textit{system under test} (SUT), namely, white-box testing~\cite{ostrand2002white}, gray-box testing~\cite{linzhang2004generating} and black-box testing~\cite{edwards2001framework}. These methods are hereby described:
\begin{description}[align=left]
	\item[White-box Testing] The internals of the SUT are all visible and known, and, as such, this information can be used to create test scenarios. Additionally, white-box testing is not restricted to failure detection but is also able to detect errors.
	\item[Black-box Testing] The SUT internal content is hidden, and the only knowledge about the system's or module's inputs and outputs is known, being closer to real-world use situations.
	\item[Gray-box Testing] A mix of the two previous techniques is used. Information about the internals of the SUT is used, but, however, tests are conducted under realistic conditions, where only failures are detected.
\end{description}


IoT systems are complex by nature, depending on different software and hardware components, modules and architectures, produced by many manufacturers and with different working properties. As such, diverse needs of testing appear in the result of the different variables that need to be tested. One can identify various challenges, as, for example, the high-heterogeneity, large-scale, dynamic environment, real-time needs, security and privacy implications, and the difficulty with test automation. Hence, different testing needs appear from the different IoT layers (Figure \ref{fig:iotlayers}):
\begin{description}[align=left]
	\item[Edge Testing]: Concerns the testing of the more low-level parts of IoT system's, like micro-controllers (e.g. Arduino) and programmable logic controllers (PLC). Testing approaches like embedded system testing can be typically used to perform tests on the edge layer, asserting the edge devices against their specification~\cite{koopman2011embedded}.
	\item[Fog Testing]: Tests regarding the middle-point layer on IoT system's, normally composed of gateways. Software testing approaches can be seamlessly applied since the devices that belong to this layer have, typically, a comfortable amount of computing power and memory, running full operating systems (e.g. Linux). Additionally, since this is the connectivity-enabler layer, connecting the devices and the Internet \textit{per se}, it should cover network testing~\cite{Kirichek2016} and security testing~\cite{6978614}.
	\item[Cloud Testing]: Cloud testing addresses the need of testing the unique quality concerns of the cloud infrastructure such as massive scalability and dynamic configuration. This field has open-challenges and issues of its own, and they are extensively analysed in the literature~\cite{bai2011cloud,riungu2010research}.
\end{description}

To be able to test IoT systems as a whole, work has been pursued towards \textit{IoT testbeds}, that enable to test the IoT systems from lower layers until the high-level ones. Although almost every testbed vertically encompasses all the layers, they are \textit{single-domain}, focusing on a specific domain of application or technological aspect. Although there are some \textit{multi-domain} testbeds that combine different technologies into a common experimental facility. A survey on the currently active and publicly available physical testbeds is given by Gluhak et al.~\cite{gluhak2011survey}.

Along with physical testbeds, another approach that has been pursued testing IoT-based systems is the use of emulators and simulators. On one hand, an emulator is a system that behaves exactly like the target system (e.g. physical devices emulation). On the other hand, simulators enable a close replication of the target system but implemented in an entirely different way (e.g. smart city simulation). The work pursued by Looga et al. surveys the existent simulators and emulators, revealing issues on their suitability for testing IoT-based system and proposing a new emulation platform for the IoT (MAMMotH)~\cite{looga2012mammoth}.

\subsection{Internet-of-Things Testing Solutions} \label{tools}

As of today, there are already some solutions available for testing IoT-based systems. These solutions focus on different IoT layers and enabling technologies. The tools have been selected after a curated search on scholar database (Scopus) and Google search engine, and are shortly described in the following paragraphs. The keywords used for the search were one or a combination of the following: \textsc{Internet-of-Things}, \textsc{test}, \textsc{testing} and \textsc{IoT}.

\begin{description}[align=left]
	\item[PlatformIO] \url{http://platformio.org/}
	
	PlatformIO is a cross-platform code builder and library manager, supporting nearly 200 development boards and most major embedded software development platforms. It has a unit testing feature (PIO Unit Testing) which is based on the Unity Test API by \textit{ThrowTheSwitch.org}~\cite{unity}.
	
	\item[IoTIFY] \url{https://iotify.io/}
	
	IoTIFY is an application development environment for IoT without hardware dependencies. By resorting to device virtualization, it provides a virtual lab for building embedded prototypes and a network simulation for system scaling and data generation. 
	
	\item[FIT IoT-LAB] \url{https://www.iot-lab.info/}\cite{adjih2015fit}
	
	IoT-LAB is a scientific testbed for testing small wireless sensor devices and heterogeneous communicating objects built on a very large scale infrastructure, deployed around six sites in France with over 2000 sensor nodes. It is the successor of SENSLAB testbed and is part of the \textit{Future Internet of the Things} (FIT) platform. 
	
	\item[ArduinoUnit] \url{https://github.com/mmurdoch/arduinounit}
	
	ArduinoUnit is a unit testing framework for Arduino libraries. Being a lightweight library, developers can easily test their systems in an Arduino board, despite their low amount of resources. However, it is up to the developer to upload the testing application to the target board and the results must be interpreted by them, commonly through the use of a serial port monitor.
	
	\item[MAMMotH]~\cite{looga2012mammoth}
	
	MAMMotH is a large-scale IoT emulator, being able to emulate ten thousand devices per Virtual Machine (VM), whose architecture presumes three distinct scenarios, namely: mobile devices connected via GPRS to a base station forming a star topology, a stand-alone wireless sensor network (WSN) connected to a base station via GPRS and constrained devices (e.g. sensors) connect to proxies, which in turn connect to the backend, a large-scale IoT emulator.
	In order to reproduce the communication problems present in a real IoT environment, the proxy to which the devices are connected simulates a radio link for each node, able to delay and drop messages. Developers can then use this setup to create experiment scenarios, deploy them on a testbed and monitor the results.
	
	\item[SimIoT]~\cite{Sotiriadis2014}
	
	SimIoT is a toolkit to achieve experimentation on dynamic and real-time multi-user submissions within an IoT scenario. The toolkit is based on the SimIC, a system that allows modelers to configure a diversity of clouds in terms of datacenter hosts and software policies wherein the desired number of users could send single or multiple requests for computational power, software resources, and duration of VM virtualization.
	
	\item[Cooja Simulator] \url{https://anrg.usc.edu/contiki/}~\cite{bagula2015iot}
	
	The Cooja Simulator is an emulation/simulation platform developed for the Contiki OS. It is an extensible Java-based simulator able to simulate the network, operating system, and instruction set. It is also able to emulate the execution of the exact same firmware that may be uploaded to physical nodes, instead of simulating it. Cooja allows developers to test their code and systems long before running it on the target hardware.
	
	\item[TOSSIM] \url{http://tinyos.stanford.edu/} \cite{levis2003tossim}
	
	TOSSIM is a wireless sensor network simulator that was built with the specific goal to simulate TinyOS devices. Since TinyOS is event-based, it is easily translated into a simulator engine with discrete events, thus simplifying it and making it more effective. TOSSIM supports two programming interfaces (Python, C++), and has various levels of simulation, from hardware interrupts to high-level system events, such as packet arrivals.
	
	\item[iFogSim] \url{http://www.cloudbus.org/cloudsim/}
	
	iFogSim is a Fog Computing Simulator able to simulate edge devices, cloud data centers, and network links, and perform metrics evaluation on them. With these features, it allows investigation and comparison of resource management techniques based on QoS (Quality-of-Service) criteria (e.g. latency, network congestion).

	\item[MobIoTSim] \url{https://github.com/sed-szeged/MobIoTSim} \cite{Pflanzner2016}
	
	MobIoTSim is a mobile IoT device simulator, developed in Android, designed to help researchers learn IoT device handling without buying real sensors, and to test and demonstrate IoT applications utilizing multiple devices. This system can be connected to a gateway service in a cloud, such as IBM Bluemix Platform and Azure IoT Hub, to manage the simulated devices and to send back notifications by responding to critical sensor values. By using this tool, developers can examine the behavior of small IoT systems, and evaluate IoT cloud applications with a hand-held device.
	
	\item[IOTSim] \cite{Zeng2017}
	
	IOTSim is a Cloud simulator built on top of the CloudSim system and designed to support the testing of IoT big data processing, resorting to a MapReduce approach. By inherently supporting big data systems, it facilitates the understanding and analysis of the impact and performance of IoT-based applications by researchers and commercial organizations.

	\item[DPWSim] \cite{Han2014}
	
	DPWSim is a simulation toolkit to support the prototyping and development of service-oriented and event-driven IoT applications. It aims to support the OASIS standard Devices Profile for Web Services (DPWS), which, although it enables the use of web services on smart and resource-constrained devices, reduces the scope of such a system to IoT devices that implement the referred device profile.
	
	\item[SimpleIoTSimulator] \url{https://www.smplsft.com/SimpleIoTSimulator.html}
	
	SimpleIoTSimulator is an IoT device simulator that can create test environments comprised of thousands of sensors on a single computer. It supports many common IoT protocols and is able to learn from data of recorded packet exchanges from real servers and sensors and model the behavior of its simulated devices from such data.
	
	\item[Atomiton IoT Simulator] \url{http://www.atomiton.com/}
	
	The Atomiton IoT Simulator, built atop Atomiton Stack (a proprietary operating environment for the Internet-of-Things), is a prototyping and testing framework able to simulate virtual sensors, actuators, and devices with unique behaviors. It allows prototyping an IoT solution and tests its scalability by providing the ability to create boundary test cases, resorting to the simulation of thousands of devices and events such as network interruptions, device response delays, and peak load.
	
	\item[MBTAAS] \cite{Ahmad2016}
	
	The Model-Based Testing as a Service (MBTAAS) allows the systematically test the IoT and data platforms. The approach resorts to a combination of model-based testing (MBT) techniques and service-oriented solutions. The solution has been tested on top of the FIWARE IoT-enabling platform. Further, the modularity of the solutions allows \textit{integration testing} between different IoT platforms.
	
	\item[CupCarbon] \url{http://www.cupcarbon.com/} \cite{Bounceur2016}
	
	CupCarbon is a platform for designing smart-city and IoT Wireless Sensor Networks (SCI-WSN). It is designed around two simulation environments, namely: one model's mobile units (e.g. cars) and natural events (e.g. wildfire, gas) and the other makes discrete event simulation of wireless sensor networks and is able to take into account the scenario designed in the previous environment. With the integration of the OpenStreeMaps framework and allowing the programming of each node individually, CupCarbon is a useful tool to design, visualize, debug and validate distributed algorithms for monitoring and environmental data collection.
\end{description}

\begin{table}
	\centering
	\begin{adjustbox}{width=\textwidth,max totalheight=\textheight}
		\begin{tabular}{P{1.6cm} P{1.2cm} P{1.7cm} l P{1.6cm} P{1.2cm} P{1.6cm} P{1.3cm} P{1.5cm} P{1.45cm} P{1.2cm}}
			\toprule
			\textbf{Tool}  & \textbf{IoT Layer} & \textbf{Test Level} & \textbf{Test Method} & \textbf{Testing Artifact} & \textbf{Prog. Lang.} & \textbf{Test Environment} & \textbf{Test Runner} & \textbf{Sup. Platforms} & \textbf{Scope}   & \textbf{License} \\ 
			\hline
			\textbf{PlatformIO}             & Edge       & Unit                & White-box   & Code                          & C/C++, Arduino    & Device                & Local , Remote & 15+            & Market           & Close  \\ \midrule
			\textbf{IoTIFY}                 & All        & Any                 & White-box   & N/A                           & N/A                 & Simulator        & Remote         & N/A              & Market           & Close  \\ \midrule
			\textbf{FIT IoT-LAB}            & All        & Any                 & Any         & N/A                          & N/A                 & Physical Testbed & Local, Remote  & 6+             & Academic, Market & Open    \\ \midrule
			\textbf{ArduinoUnit}            & Edge       & Unit                & White-box   & Code                          & Arduino           & Device       & Local          & Arduino        & Academic, Market & Open    \\ \midrule
			\textbf{MAMMotH}                & All        & Integration, System & Any         & Network          & N/A                 & Emulator         & Local          & N/A              & Academic         & N/A       \\ \midrule
			\textbf{Cooja}                  & Edge       & Integration         & Black-box   & Network                       & C                 & Emulator         & Local          & Contiki OS     & Academic, Market & N/A       \\ \midrule
			\textbf{TOSSIM}                 & Edge       & Integration         & Any         & Application, Network          & Python, C++       & Simulator        & Local          & TinyOS         & Academic         & Open    \\ \midrule
			\textbf{SWE Simulator}          & Edge       & System              & Black-box   & Application, Network          & XML, Visual       & Simulator        & Local          & SWE Standard   & Academic         & N/A       \\ \midrule
			\textbf{SimIoT}                 & Fog        & Integration, System & Black-box   & Any                           & N/A                 & Simulator        & Local          & N/A              & Academic         & N/A       \\ \midrule
			\textbf{iFogSim}                & Edge, Fog  & Integration, System & Grey-box    & Network                       & Java              & Simulator        & Local          & N/A              & Academic         & Open    \\ \midrule
			\textbf{MobIoTSim}              & Fog, Cloud & Integration, System & Grey-box    & Application, Network           & N/A                 & Simulator        & Local          & N/A              & Academic         & Open    \\ \midrule
			\textbf{IOTSim}                 & Cloud      & Integration         & Any         & Application  & N/A                 & Simulator        & N/A              & N/A             & Academic         & N/A       \\ \midrule
			\textbf{DPWSim}                 & Fog, Cloud & Integration, System & Any         & Application                   & WSDL              & Simulator        & Local          & DPWS           & Academic         & N/A       \\ \midrule
			\textbf{SimpleIoTSimulator}     & Edge, Fog  & Integration, System                 & Any         & Network                       & N/A                & Simulator        & Local          & N/A              & Market           & Close  \\ \midrule
			\textbf{Atomiton IoT Simulator} & All        & Any                 & Grey-box    & N/A                         & N/A                 & Simulator        & Remote         & N/A             & Market           & Close  \\ \midrule
			\textbf{MBTAAS}              & All        & Any              & Black-box         & Model                       & OCL         & Platform        & N/A          & N/A             & Academic         & N/A   \\\bottomrule
			\textbf{CupCarbon}              & All        & System              & Any         & Network                       & SenScript         & Simulator        & Local          & -              & Academic         & Open   \\\bottomrule
		\end{tabular}
	\end{adjustbox}
	\caption[Overview comparison  of the available tools for IoT testing]{Overview comparison  of the available tools on the IoT testing landscape. N/A symbolizes that there is no information available or not have been found during our research.}
	\label{tab:overview}
\end{table}


An overview comparison of the available tools for testing IoT solutions is given in Table \ref{tab:overview}. Testing capabilities of each solution are analyzed by the observation of different variables. 

In the first place, the tools are divided by the \textit{IoT Layer} they focus on, as they are presented in Figure \ref{fig:iotlayers}. Here it can be observed a relation between the layer and the testing variable related to. Edge layer tools, such as the \textsc{PlatformIO} and \textsc{ArduinoUnit}, typically focus on testing the code that runs on edge devices (e.g. Arduino). However, to test the edge layer the already available tools from embedded system testing can be helpful (e.g. \textsc{UNITY}\footnote{UNITY by ThrowTheSwitch.org: \url{https://www.throwtheswitch.org/unity}}). Fog and Cloud tools are typically concerned about network or application testing, disregarding the low-level tests on code but testing at the System and Integration level. 

By the analysis of the \textit{Test Level} to which each tool is concerned about, we notice that there are tools covering all levels, from \textit{unit testing} until \textit{acceptance testing}, at least in a partial way. We must note that although some tools enable one to test all the levels, they do not provide out-of-the-box functionalities to do so. Example of one of those is the \textsc{FIT IoT-LAB} testbed that provides a large-scale platform to test applications across the different layers, but requires development efforts in, for example, retrieve and manage data from that testing. In other cases, the tools provide only partial support for the testing functionalities, e.g., providing functionalities of collecting all network logs and responses but not providing direct insights on that information.

Some gaps appear in the solutions support of different languages and platforms. A vast part of the available tools focuses on a specific platform, language or standard, lacking the support for the heterogeneity of the IoT field. Example of such tools are the \textsc{DPWSim} that focus on the Devices Profile for Web Services (DPWS) standard language and the \textsc{TOSSIM} simulator for the \textit{TinyOS} compatible devices. Another problem appears from the large range of network communication protocols and IoT-enabling technologies (e.g. reference architectures) that are now appearing in the market without any kind of standardization, which leads to the lack of tools to test them in a platform-agnostic way. However, some have a large number of supported platforms or are open to any implementation by requiring some extra development efforts.

Moving towards the different artifacts that need to be tested in the IoT landscape, the testing necessities are common to the highly-distributed systems field. Firstly, and the artifact more covered by the available solutions (e.g. \textsc{MAMMotH}, \textsc{iFogSim}), the network and communication variable. Secondly, with some available tools such as the \textsc{MobIoTSim}, the application level testing, in which the functionality, usability, and consistency can be tested within a real-world scenario, disregarding the business logic behind them. Some solutions are also available for code testing for the edge devices such as the \textsc{PlatformIO}. However, it is easily noticeable that there is a lack of tools for testing certain artifacts such as security and privacy, regulatory testing and firmware/software upgrade (e.g. out-of-the-box continuous integration functionalities). 

In the security and privacy scope, there is work being pursued by the OWASP (Open Web Application Security Project) to ``help manufacturers, developers, and consumers better understand the security issues associated with the Internet-of-Things, and to enable users in any context to make better security decisions when building, deploying, or assessing IoT technologies'' \cite{owasp}. 

Testing environments are another distinguishable aspect within the available testing solutions. Most of the environments are purely virtual by means of emulation (e.g. virtual representation of an Arduino board) or simulation techniques (e.g. simulation of a smart city or smart house). However, some efforts have been done in the creation of physical testbeds like the \textsc{FIT IoT-LAB}. Also, some traditional software testing tools are available (for unit testing purposes) that most of the times rely on physical devices to conduct the testing. 

Another relevant aspect is the stage of development of the solutions found and their openness. It is observable that most of the solutions have been presented in the literature, however, most of them are purely academic and there is no access to its source code or the software package. Comparatively, the solutions available to be used are scarce and most of them are closed-source, reducing the possibility of extending the tool functionalities or improving it by the means of extensions or plug-ins. Here it can also be noted that some tools are only available on remote test runners which can reduce the ability to test specific needs of certain solutions and raise privacy concerns.

\subsection{Conclusions} \label{conclusions}



As presented in the previous sections, testing techniques and methodologies have long been developed and studied across software and hardware study areas independently but also, though with less impact, jointly. The emergence of the IoT as a highly market-valuable area, with a wide range of application scenarios, has intensified the needs, and consequently the efforts, in testing high-scale solutions based both in software and hardware.

However, due to the cross-domain particularities of the IoT, long-pursued and pending research challenges from other study areas are now also becoming a problem in the IoT field. One can enumerate the following areas as the ones with larger significance:
\begin{enumerate}
	\item \textit{Heterogeneous Systems}: Testing heterogeneous systems' challenges appears from the integration and system-level testing perspective. Although there are some techniques such as \textit{Manual Exploratory Testing}, \textit{Combinatorial Testing} and \textit{Search-Based Software Testing}, there are still a considerable number of gaps, resulting in part from differences in industry focus and research focus~\cite{Ghazi2015}.
	\item \textit{Large-Scale Distributed Systems}: Large-scale and highly-distributed systems lead to the appearance of new variables that need to be tested being some of them still open issues on the literature. Examples are the load testing~\cite{7123673} and the handling of the dynamic behavior of such systems.
	\item \textit{Cloud-based Systems}: The high-level layer of IoT systems is the cloud. Although cloud computing has become ubiquitous nowadays, there are still gaps in how to test cloud-based/cloud-connected systems. Example of such is the design and testing of elastic cloud-based solutions~\cite{calheiros2011cloudsim}.
	\item \textit{Embedded Software Systems}: Devices typically have constraints of memory and processing power, which make it hard for testing the software running on them. Also, this kind of devices are typically associated with real-time needs and are prone to fail due to hardware problems (e.g. power surge) which makes the testing responses more volatile to environmental changes~\cite{BANERJEE2016121}.
\end{enumerate}

Even further, we can consider that there is a gap on tools that can test distributed and heterogeneous systems, especially in an automated way, however, there exists ongoing work on covering these lacks~\cite{Lima2016}.

Combined with these open gaps in the literature, spread among the technologies that empower the IoT-based solutions, there is a lack of a proper solution or methodology for testing IoT solutions with the already spread knowledge about testing in software and in hardware. The pursued aggregation and comparison of the available solutions for testing purposes shows that there is just a small number of solutions that test IoT systems across all the layers, and those solutions have always limitations of some sort (e.g. limited number of testing variables or supported platforms).

As such, research focus should be given in the enhancement of available tools for carrying tests on IoT and its layers, under different scales and criticality levels, without disregarding the existing and continuously growing heterogeneity in devices, communication protocols, standards, and reference architectures. Nonetheless, tools with different testing focus should be developed. 

Furthermore, testing solutions should be designed taking into account test automation needs and continuous integration functionalities, in order to enable the implementation of testing and deployment pipelines for IoT solutions.


The key features that differentiate IoT testing needs from the traditional systems are the heterogeneous and large-scale objects and networks. These factors lead to an increase in the complexity and difficulty of testing IoT-based solutions. As such, this article addresses the actual state-of-the-art techniques and methodology largely widespread in the software development community and the need for bringing such techniques and methodology to the IoT development scope. Then, the gaps in the currently available testing solutions are aggregated by an inspection of the commercial and open source tools for testing such systems. Within this, we consider that there is a set of old-known challenges that is now influencing directly the IoT systems and that further work must be pursued on the development of testing solutions, automation procedures for testing and continuous integration features.

\section{Conclusions}

With the advent of the Internet-of-Things, the vision Internet has shifted beyond the World Wide Web (WWW) to a world of connected things, ranging from sensors to actuators, integrated into a variety of daily objects and infrastructures. This reality shift, empowered by emergent network technologies plus cheap and low-powered devices, has opened a wide range of opportunities both from research and business perspectives, in fields such as industry, city management, and (e)health.

IoT has already a noticeable impact on our daily-lives, however, several challenges are still open from conception to the maintenance of such systems. In an era of cloud-first design, IoT reality has increased the interest in new architectural approaches as it is observable by the birth of fog and mist/edge computing.

In terms of software engineering body-of-knowledge, approaches such as Visual Programming and Model-driven (Software) Engineering have for long being used to tackle the complexity of software systems, as it is the case of IoT. More recently, work has been pursued on live programming and the use of it as the way to increase the liveness of software systems development, improving the developer overall experience (e.g. reducing the complexity of detecting and correct bugs).



\bibliographystyle{ACM-Reference-Format}
\bibliography{bibliography}


\begin{thebibliography}{96}


\ifx \showCODEN    \undefined \def \showCODEN     #1{\unskip}     \fi
\ifx \showDOI      \undefined \def \showDOI       #1{#1}\fi
\ifx \showISBNx    \undefined \def \showISBNx     #1{\unskip}     \fi
\ifx \showISBNxiii \undefined \def \showISBNxiii  #1{\unskip}     \fi
\ifx \showISSN     \undefined \def \showISSN      #1{\unskip}     \fi
\ifx \showLCCN     \undefined \def \showLCCN      #1{\unskip}     \fi
\ifx \shownote     \undefined \def \shownote      #1{#1}          \fi
\ifx \showarticletitle \undefined \def \showarticletitle #1{#1}   \fi
\ifx \showURL      \undefined \def \showURL       {\relax}        \fi
\providecommand\bibfield[2]{#2}
\providecommand\bibinfo[2]{#2}
\providecommand\natexlab[1]{#1}
\providecommand\showeprint[2][]{arXiv:#2}

\bibitem[\protect\citeauthoryear{??}{Fee}{2016}]%
        {Feeney2016}
 \bibinfo{year}{2016}\natexlab{}.
\newblock \bibinfo{title}{{A Primer on Continuous Delivery}}.
\newblock
\newblock
\urldef\tempurl%
\url{https://feeney.mba/2016/02/25/a-primer-on-continuous-delivery}
\showURL{%
\tempurl}
\newblock
\shownote{[Online; accessed 1. Jun. 2018].}


\bibitem[\protect\citeauthoryear{??}{Int}{2017}]%
        {IntSta17}
 \bibinfo{year}{2017}\natexlab{}.
\newblock \bibinfo{booktitle}{\emph{{Global Internet Report 2017: Paths to Our
  Digital Future}}}.
\newblock \bibinfo{type}{{T}echnical {R}eport}. \bibinfo{institution}{Internet
  Society}. \bibinfo{pages}{148} pages.
\newblock


\bibitem[\protect\citeauthoryear{??}{cis}{2018}]%
        {ciscoioe}
 \bibinfo{year}{2018}\natexlab{}.
\newblock \bibinfo{title}{{Internet of Everything (IoE)}}.
\newblock
\newblock
\urldef\tempurl%
\url{https://newsroom.cisco.com/ioe}
\showURL{%
\tempurl}
\newblock
\shownote{[Online; accessed 14. May 2018].}


\bibitem[\protect\citeauthoryear{??}{OWA}{2018}]%
        {OWASP2018Apr}
 \bibinfo{year}{2018}\natexlab{}.
\newblock \bibinfo{title}{{OWASP Internet of Things Project}}.
\newblock
\newblock
\urldef\tempurl%
\url{https://www.owasp.org/index.php/OWASP_Internet_of_Things_Project#IoT_Attack_Surface_Areas_Project}
\showURL{%
\tempurl}
\newblock
\shownote{[Online; accessed 22. May 2018].}


\bibitem[\protect\citeauthoryear{Adjih, Baccelli, Fleury, Harter, Mitton, Noel,
  Pissard-Gibollet, Saint-Marcel, Schreiner, Vandaele, et~al\mbox{.}}{Adjih
  et~al\mbox{.}}{2015}]%
        {adjih2015fit}
\bibfield{author}{\bibinfo{person}{Cedric Adjih}, \bibinfo{person}{Emmanuel
  Baccelli}, \bibinfo{person}{Eric Fleury}, \bibinfo{person}{Gaetan Harter},
  \bibinfo{person}{Nathalie Mitton}, \bibinfo{person}{Thomas Noel},
  \bibinfo{person}{Roger Pissard-Gibollet}, \bibinfo{person}{Frederic
  Saint-Marcel}, \bibinfo{person}{Guillaume Schreiner}, \bibinfo{person}{Julien
  Vandaele}, {et~al\mbox{.}}} \bibinfo{year}{2015}\natexlab{}.
\newblock \showarticletitle{FIT IoT-LAB: A large scale open experimental IoT
  testbed}. In \bibinfo{booktitle}{\emph{IEEE 2nd World Forum on Internet of
  Things}}. IEEE, \bibinfo{pages}{459--464}.
\newblock


\bibitem[\protect\citeauthoryear{Ahmad, Bouquet, Fourneret, Le~Gall, and
  Legeard}{Ahmad et~al\mbox{.}}{2016}]%
        {Ahmad2016}
\bibfield{author}{\bibinfo{person}{Abbas Ahmad}, \bibinfo{person}{Fabrice
  Bouquet}, \bibinfo{person}{Elizabeta Fourneret}, \bibinfo{person}{Franck
  Le~Gall}, {and} \bibinfo{person}{Bruno Legeard}.}
  \bibinfo{year}{2016}\natexlab{}.
\newblock \bibinfo{booktitle}{\emph{Model-Based Testing as a Service for IoT
  Platforms}}.
\newblock \bibinfo{publisher}{Springer International Publishing},
  \bibinfo{address}{Cham}, \bibinfo{pages}{727--742}.
\newblock
\showISBNx{978-3-319-47169-3}
\urldef\tempurl%
\url{https://doi.org/10.1007/978-3-319-47169-3_55}
\showDOI{\tempurl}


\bibitem[\protect\citeauthoryear{Alexander, Ishikawa, and
  Silverstein}{Alexander et~al\mbox{.}}{1977}]%
        {Alexander1977}
\bibfield{author}{\bibinfo{person}{C Alexander}, \bibinfo{person}{S Ishikawa},
  {and} \bibinfo{person}{M Silverstein}.} \bibinfo{year}{1977}\natexlab{}.
\newblock \bibinfo{booktitle}{\emph{{A Pattern Language}}}.
\newblock
\showISBNx{13-978-0-19-501919-3}
\showISSN{01950191}
\urldef\tempurl%
\url{https://doi.org/10.2307/1574526}
\showDOI{\tempurl}


\bibitem[\protect\citeauthoryear{Bagula and Erasmus}{Bagula and
  Erasmus}{2015}]%
        {bagula2015iot}
\bibfield{author}{\bibinfo{person}{B Bagula} {and} \bibinfo{person}{ZENVILLE
  Erasmus}.} \bibinfo{year}{2015}\natexlab{}.
\newblock \showarticletitle{Iot emulation with cooja}. In
  \bibinfo{booktitle}{\emph{ICTP-IoT Workshop}}.
\newblock


\bibitem[\protect\citeauthoryear{Bai, Li, Chen, Tsai, and Gao}{Bai
  et~al\mbox{.}}{2011}]%
        {bai2011cloud}
\bibfield{author}{\bibinfo{person}{Xiaoying Bai}, \bibinfo{person}{Muyang Li},
  \bibinfo{person}{Bin Chen}, \bibinfo{person}{Wei-Tek Tsai}, {and}
  \bibinfo{person}{Jerry Gao}.} \bibinfo{year}{2011}\natexlab{}.
\newblock \showarticletitle{Cloud testing tools}. In
  \bibinfo{booktitle}{\emph{Service Oriented System Engineering (SOSE), 2011
  IEEE 6th International Symposium on}}. IEEE, \bibinfo{pages}{1--12}.
\newblock


\bibitem[\protect\citeauthoryear{Banerjee, Chattopadhyay, and
  Roychoudhury}{Banerjee et~al\mbox{.}}{2016}]%
        {BANERJEE2016121}
\bibfield{author}{\bibinfo{person}{Abhijeet Banerjee}, \bibinfo{person}{Sudipta
  Chattopadhyay}, {and} \bibinfo{person}{Abhik Roychoudhury}.}
  \bibinfo{year}{2016}\natexlab{}.
\newblock \showarticletitle{Testing Embedded Software}.
\newblock \bibinfo{series}{Advances in Computers}, Vol.~\bibinfo{volume}{101}.
  \bibinfo{publisher}{Elsevier}, \bibinfo{pages}{121 -- 153}.
\newblock
\showISSN{0065-2458}
\urldef\tempurl%
\url{https://doi.org/10.1016/bs.adcom.2015.11.005}
\showDOI{\tempurl}


\bibitem[\protect\citeauthoryear{Barricelli and Valtolina}{Barricelli and
  Valtolina}{2015}]%
        {Barricelli2015}
\bibfield{author}{\bibinfo{person}{Barbara~Rita Barricelli} {and}
  \bibinfo{person}{Stefano Valtolina}.} \bibinfo{year}{2015}\natexlab{}.
\newblock \bibinfo{booktitle}{\emph{Designing for End-User Development in the
  Internet of Things}}.
\newblock \bibinfo{publisher}{Springer International Publishing},
  \bibinfo{address}{Cham}, \bibinfo{pages}{9--24}.
\newblock
\showISBNx{978-3-319-18425-8}
\urldef\tempurl%
\url{https://doi.org/10.1007/978-3-319-18425-8_2}
\showDOI{\tempurl}


\bibitem[\protect\citeauthoryear{Beizer}{Beizer}{2003}]%
        {beizer2003software}
\bibfield{author}{\bibinfo{person}{Boris Beizer}.}
  \bibinfo{year}{2003}\natexlab{}.
\newblock \bibinfo{booktitle}{\emph{{Software testing techniques}}}.
\newblock \bibinfo{publisher}{Dreamtech Press}.
\newblock


\bibitem[\protect\citeauthoryear{Bericat}{Bericat}{2018}]%
        {redhatiotenter}
\bibfield{author}{\bibinfo{person}{David Bericat}.}
  \bibinfo{year}{2018}\natexlab{}.
\newblock \bibinfo{title}{{Internet of Things (IoT) overview powered by Red
  Hat}}.
\newblock
\newblock
\urldef\tempurl%
\url{https://www.slideshare.net/DavidBericat/ss45534bericatlacimaaccentureintro}
\showURL{%
\tempurl}
\newblock
\shownote{[Online; accessed 22. May 2018].}


\bibitem[\protect\citeauthoryear{Blackstock and Lea}{Blackstock and
  Lea}{2012a}]%
        {Blackstock2012}
\bibfield{author}{\bibinfo{person}{Michael Blackstock} {and}
  \bibinfo{person}{Rodger Lea}.} \bibinfo{year}{2012}\natexlab{a}.
\newblock \showarticletitle{{IoT mashups with the WoTKit}}.
\newblock \bibinfo{journal}{\emph{Proceedings of 2012 International Conference
  on the Internet of Things, IOT 2012}} (\bibinfo{year}{2012}),
  \bibinfo{pages}{159--166}.
\newblock
\showISBNx{9781467313469}
\urldef\tempurl%
\url{https://doi.org/10.1109/IOT.2012.6402318}
\showDOI{\tempurl}


\bibitem[\protect\citeauthoryear{Blackstock and Lea}{Blackstock and
  Lea}{2012b}]%
        {blackstock2012iot}
\bibfield{author}{\bibinfo{person}{Michael Blackstock} {and}
  \bibinfo{person}{Rodger Lea}.} \bibinfo{year}{2012}\natexlab{b}.
\newblock \showarticletitle{IoT mashups with the WoTKit}. In
  \bibinfo{booktitle}{\emph{Internet of Things (IOT), 2012 3rd International
  Conference on the}}. \bibinfo{publisher}{IEEE}, \bibinfo{pages}{159--166}.
\newblock


\bibitem[\protect\citeauthoryear{Blackstock and Lea}{Blackstock and
  Lea}{2014}]%
        {Blackstock2014}
\bibfield{author}{\bibinfo{person}{Michael Blackstock} {and}
  \bibinfo{person}{Rodger Lea}.} \bibinfo{year}{2014}\natexlab{}.
\newblock \showarticletitle{{Toward a Distributed Data Flow Platform for the
  Web of Things (Distributed Node-RED)}}. In
  \bibinfo{booktitle}{\emph{Proceedings of the 5th International Workshop on
  Web of Things - WoT '14}}. \bibinfo{pages}{34--39}.
\newblock
\showISBNx{9781450330664}
\urldef\tempurl%
\url{https://doi.org/10.1145/2684432.2684439}
\showDOI{\tempurl}


\bibitem[\protect\citeauthoryear{Boshernitsan and Downes}{Boshernitsan and
  Downes}{2004}]%
        {Boshernitsan2004}
\bibfield{author}{\bibinfo{person}{Marat Boshernitsan} {and}
  \bibinfo{person}{Michael Downes}.} \bibinfo{year}{2004}\natexlab{}.
\newblock \showarticletitle{{Visual Programming Languages: A Survey}}.
\newblock \bibinfo{journal}{\emph{Computer Science Division (EECS)}}
  \bibinfo{number}{December} (\bibinfo{year}{2004}).
\newblock
\urldef\tempurl%
\url{http://digitalassets.lib.berkeley.edu/techreports/ucb/text/CSD-04-1368.pdf}
\showURL{%
\tempurl}


\bibitem[\protect\citeauthoryear{Bounceur}{Bounceur}{2016}]%
        {Bounceur2016}
\bibfield{author}{\bibinfo{person}{Ahc{\`{e}}ne Bounceur}.}
  \bibinfo{year}{2016}\natexlab{}.
\newblock \showarticletitle{{CupCarbon: A New Platform for Designing and
  Simulating Smart-City and IoT Wireless Sensor Networks (SCI-WSN)}}.
\newblock \bibinfo{journal}{\emph{Proceedings of the International Conference
  on Internet of Things and Cloud Computing}} (\bibinfo{year}{2016}),
  \bibinfo{pages}{1:1----1:1}.
\newblock
\showISBNx{978-1-4503-4063-2}
\showISSN{978-1-4503-4063-2}
\urldef\tempurl%
\url{https://doi.org/10.1145/2896387.2900336}
\showDOI{\tempurl}


\bibitem[\protect\citeauthoryear{Burnett and Baker}{Burnett and Baker}{1994}]%
        {Burnett1994287}
\bibfield{author}{\bibinfo{person}{M.M. Burnett} {and} \bibinfo{person}{M.J.
  Baker}.} \bibinfo{year}{1994}\natexlab{}.
\newblock \showarticletitle{A classification system for visual programming
  languages}.
\newblock \bibinfo{journal}{\emph{Journal of Visual Languages and Computing}}
  \bibinfo{volume}{5}, \bibinfo{number}{3} (\bibinfo{year}{1994}),
  \bibinfo{pages}{287--300}.
\newblock
\urldef\tempurl%
\url{https://doi.org/10.1006/jvlc.1994.1015}
\showDOI{\tempurl}
\newblock
\shownote{cited By 22.}


\bibitem[\protect\citeauthoryear{Bushmann, Meunier, and Rohnert}{Bushmann
  et~al\mbox{.}}{1996}]%
        {Bushmann1996}
\bibfield{author}{\bibinfo{person}{F Bushmann}, \bibinfo{person}{R Meunier},
  {and} \bibinfo{person}{H Rohnert}.} \bibinfo{year}{1996}\natexlab{}.
\newblock \showarticletitle{{Pattern-oriented software architecture: A system
  of patterns}}.
\newblock \bibinfo{journal}{\emph{John Wiley{\&}Sons}}  \bibinfo{volume}{1}
  (\bibinfo{year}{1996}), \bibinfo{pages}{476}.
\newblock
\showISBNx{0471958697}
\showISSN{0007-1250}
\urldef\tempurl%
\url{https://doi.org/10.1192/bjp.108.452.101}
\showDOI{\tempurl}


\bibitem[\protect\citeauthoryear{Buyya and Dastjerdi}{Buyya and
  Dastjerdi}{2016}]%
        {Buyya2016}
\bibfield{author}{\bibinfo{person}{Rajkumar Buyya} {and}
  \bibinfo{person}{Amir~Vahid Dastjerdi}.} \bibinfo{year}{2016}\natexlab{}.
\newblock \bibinfo{booktitle}{\emph{{Internet of Things: Principles and
  Paradigms}}}.
\newblock \bibinfo{publisher}{Elsevier}.
\newblock
\showISBNx{9780128053959}
\urldef\tempurl%
\url{https://doi.org/10.1016/C2015-0-04135-1}
\showDOI{\tempurl}


\bibitem[\protect\citeauthoryear{Calheiros, Ranjan, Beloglazov, De~Rose, and
  Buyya}{Calheiros et~al\mbox{.}}{2011}]%
        {calheiros2011cloudsim}
\bibfield{author}{\bibinfo{person}{Rodrigo~N Calheiros}, \bibinfo{person}{Rajiv
  Ranjan}, \bibinfo{person}{Anton Beloglazov}, \bibinfo{person}{C{\'e}sar~AF
  De~Rose}, {and} \bibinfo{person}{Rajkumar Buyya}.}
  \bibinfo{year}{2011}\natexlab{}.
\newblock \showarticletitle{CloudSim: a toolkit for modeling and simulation of
  cloud computing environments and evaluation of resource provisioning
  algorithms}.
\newblock \bibinfo{journal}{\emph{Software: Practice and experience}}
  \bibinfo{volume}{41}, \bibinfo{number}{1} (\bibinfo{year}{2011}),
  \bibinfo{pages}{23--50}.
\newblock


\bibitem[\protect\citeauthoryear{Chen and De~Luca}{Chen and De~Luca}{2016}]%
        {chen2016viple}
\bibfield{author}{\bibinfo{person}{Yinong Chen} {and} \bibinfo{person}{Gennaro
  De~Luca}.} \bibinfo{year}{2016}\natexlab{}.
\newblock \showarticletitle{VIPLE: visual IoT/robotics programming language
  environment for computer science education}. In
  \bibinfo{booktitle}{\emph{Parallel and Distributed Processing Symposium
  Workshops, 2016 IEEE International}}. \bibinfo{publisher}{IEEE},
  \bibinfo{pages}{963--971}.
\newblock


\bibitem[\protect\citeauthoryear{Cox}{Cox}{2007}]%
        {vplbook}
\bibfield{author}{\bibinfo{person}{Philip~T. Cox}.}
  \bibinfo{year}{2007}\natexlab{}.
\newblock \bibinfo{booktitle}{\emph{Visual Programming Languages}}.
\newblock \bibinfo{publisher}{John Wiley \& Sons, Inc.}
\newblock
\showISBNx{9780470050118}
\urldef\tempurl%
\url{https://doi.org/10.1002/9780470050118.ecse450}
\showDOI{\tempurl}


\bibitem[\protect\citeauthoryear{Dias, Couto, Paiva, and Ferreira}{Dias
  et~al\mbox{.}}{2018}]%
        {VVioT2018}
\bibfield{author}{\bibinfo{person}{Jo{\~{a}}o~Pedro Dias},
  \bibinfo{person}{Flavio Couto}, \bibinfo{person}{Ana C.~R. Paiva}, {and}
  \bibinfo{person}{Hugo~Sereno Ferreira}.} \bibinfo{year}{2018}\natexlab{}.
\newblock \showarticletitle{{A Brief Overview of Existing Tools for Testing the
  Internet-of-Things}}. In \bibinfo{booktitle}{\emph{Proceedings of the 2018
  IEEE International Conference on Software Testing, Verification and
  Validation Workshops (First International Workshop on Verification and
  Validation of Internet of Things)}}. \bibinfo{publisher}{IEEE - Institute of
  Electrical and Electronics Engineers}.
\newblock


\bibitem[\protect\citeauthoryear{Dias, Pinto, and Cruz}{Dias
  et~al\mbox{.}}{2017}]%
        {Dias2017}
\bibfield{author}{\bibinfo{person}{Jo{\~{a}}o~Pedro Dias},
  \bibinfo{person}{Jos{\'{e}}~Pedro Pinto}, {and}
  \bibinfo{person}{Jos{\'{e}}~Magalh{\~{a}}es Cruz}.}
  \bibinfo{year}{2017}\natexlab{}.
\newblock \showarticletitle{{A Hands-on Approach on Botnets for Behavior
  Exploration}}. In \bibinfo{booktitle}{\emph{Proceedings of the 2nd
  International Conference on Internet of Things, Big Data and Security}}.
  \bibinfo{publisher}{SCITEPRESS - Science and Technology Publications},
  \bibinfo{pages}{463--469}.
\newblock
\showISBNx{978-989-758-245-5}
\urldef\tempurl%
\url{https://doi.org/10.5220/0006392404630469}
\showDOI{\tempurl}


\bibitem[\protect\citeauthoryear{Diehl}{Diehl}{2007}]%
        {book:55167}
\bibfield{author}{\bibinfo{person}{Stephan Diehl}.}
  \bibinfo{year}{2007}\natexlab{}.
\newblock \bibinfo{booktitle}{\emph{Software Visualization: Visualizing the
  Structure, Behaviour, and Evolution of Software}}.
\newblock \bibinfo{publisher}{Springer}.
\newblock
\showISBNx{3540465049,978-3-540-46504-1}
\urldef\tempurl%
\url{http://gen.lib.rus.ec/book/index.php?md5=D6C642DE2B9229610116A0B600F02A60}
\showURL{%
\tempurl}


\bibitem[\protect\citeauthoryear{Dohr, Modre-Opsrian, Drobics, Hayn, and
  Schreier}{Dohr et~al\mbox{.}}{2010}]%
        {Dohr2010}
\bibfield{author}{\bibinfo{person}{A. Dohr}, \bibinfo{person}{R.
  Modre-Opsrian}, \bibinfo{person}{M. Drobics}, \bibinfo{person}{D. Hayn},
  {and} \bibinfo{person}{G. Schreier}.} \bibinfo{year}{2010}\natexlab{}.
\newblock \showarticletitle{{The Internet of Things for Ambient Assisted
  Living}}.
\newblock \bibinfo{journal}{\emph{Seventh International Conference on
  Information Technology: New Generations}} (\bibinfo{year}{2010}),
  \bibinfo{pages}{804--809}.
\newblock
\showISBNx{9781424462704}
\urldef\tempurl%
\url{https://doi.org/10.1109/ITNG.2010.104}
\showDOI{\tempurl}


\bibitem[\protect\citeauthoryear{Edwards}{Edwards}{2001}]%
        {edwards2001framework}
\bibfield{author}{\bibinfo{person}{Stephen~H Edwards}.}
  \bibinfo{year}{2001}\natexlab{}.
\newblock \showarticletitle{{A framework for practical, automated black-box
  testing of component-based software}}.
\newblock \bibinfo{journal}{\emph{Software Testing, Verification and
  Reliability}} \bibinfo{volume}{11}, \bibinfo{number}{2}
  (\bibinfo{year}{2001}), \bibinfo{pages}{97--111}.
\newblock


\bibitem[\protect\citeauthoryear{Ferreira}{Ferreira}{2011}]%
        {Ferreira}
\bibfield{author}{\bibinfo{person}{Hugo~Sereno Ferreira}.}
  \bibinfo{year}{2011}\natexlab{}.
\newblock \emph{\bibinfo{title}{{Adaptive Object-Modeling Patterns, Tools and
  Applications}}}.
\newblock \bibinfo{thesistype}{Ph.D. Dissertation}.
\newblock
\showISBNx{9789899689909}


\bibitem[\protect\citeauthoryear{Fleurey and Morin}{Fleurey and Morin}{2017}]%
        {7958482}
\bibfield{author}{\bibinfo{person}{F. Fleurey} {and} \bibinfo{person}{B.
  Morin}.} \bibinfo{year}{2017}\natexlab{}.
\newblock \showarticletitle{ThingML: A Generative Approach to Engineer
  Heterogeneous and Distributed Systems}.
\newblock  (\bibinfo{date}{April} \bibinfo{year}{2017}),
  \bibinfo{pages}{185--188}.
\newblock
\urldef\tempurl%
\url{https://doi.org/10.1109/ICSAW.2017.63}
\showDOI{\tempurl}


\bibitem[\protect\citeauthoryear{Francis}{Francis}{2017}]%
        {webmoz}
\bibfield{author}{\bibinfo{person}{B. Francis}.}
  \bibinfo{year}{2017}\natexlab{}.
\newblock \bibinfo{booktitle}{\emph{Web Thing API}}.
\newblock \bibinfo{type}{{T}echnical {R}eport}. \bibinfo{institution}{Mozilla}.
\newblock


\bibitem[\protect\citeauthoryear{Friedemann and Floerkemeir}{Friedemann and
  Floerkemeir}{2011}]%
        {Friedemann2011}
\bibfield{author}{\bibinfo{person}{Mattern Friedemann} {and}
  \bibinfo{person}{Christian Floerkemeir}.} \bibinfo{year}{2011}\natexlab{}.
\newblock \showarticletitle{{From the Internet to the Internet of Things}}.
\newblock \bibinfo{journal}{\emph{From Active Data Management to Event-Based
  Systems and More}} (\bibinfo{year}{2011}), \bibinfo{pages}{242--259}.
\newblock
\showISBNx{9783642238338}
\showISSN{0302-9743}
\urldef\tempurl%
\url{https://doi.org/10.1007/978-3-642-17226-7}
\showDOI{\tempurl}
\showeprint[arxiv]{9780201398298}


\bibitem[\protect\citeauthoryear{Ghazi, Petersen, and B{\"o}rstler}{Ghazi
  et~al\mbox{.}}{2015}]%
        {Ghazi2015}
\bibfield{author}{\bibinfo{person}{Ahmad~Nauman Ghazi}, \bibinfo{person}{Kai
  Petersen}, {and} \bibinfo{person}{J{\"u}rgen B{\"o}rstler}.}
  \bibinfo{year}{2015}\natexlab{}.
\newblock \bibinfo{booktitle}{\emph{Heterogeneous Systems Testing Techniques:
  An Exploratory Survey}}.
\newblock \bibinfo{publisher}{Springer International Publishing},
  \bibinfo{address}{Cham}, \bibinfo{pages}{67--85}.
\newblock
\showISBNx{978-3-319-13251-8}
\urldef\tempurl%
\url{https://doi.org/10.1007/978-3-319-13251-8_5}
\showDOI{\tempurl}


\bibitem[\protect\citeauthoryear{Gluhak, Krco, Nati, Pfisterer, Mitton, and
  Razafindralambo}{Gluhak et~al\mbox{.}}{2011}]%
        {gluhak2011survey}
\bibfield{author}{\bibinfo{person}{Alexander Gluhak}, \bibinfo{person}{Srdjan
  Krco}, \bibinfo{person}{Michele Nati}, \bibinfo{person}{Dennis Pfisterer},
  \bibinfo{person}{Nathalie Mitton}, {and} \bibinfo{person}{Tahiry
  Razafindralambo}.} \bibinfo{year}{2011}\natexlab{}.
\newblock \showarticletitle{A survey on facilities for experimental internet of
  things research}.
\newblock \bibinfo{journal}{\emph{IEEE Communications Magazine}}
  \bibinfo{volume}{49}, \bibinfo{number}{11} (\bibinfo{year}{2011}).
\newblock


\bibitem[\protect\citeauthoryear{Gubbi, Buyya, Marusic, and Palaniswami}{Gubbi
  et~al\mbox{.}}{2013}]%
        {Gubbi2013}
\bibfield{author}{\bibinfo{person}{Jayavardhana Gubbi},
  \bibinfo{person}{Rajkumar Buyya}, \bibinfo{person}{Slaven Marusic}, {and}
  \bibinfo{person}{Marimuthu Palaniswami}.} \bibinfo{year}{2013}\natexlab{}.
\newblock \showarticletitle{{Internet of Things (IoT): A vision, architectural
  elements, and future directions}}.
\newblock \bibinfo{journal}{\emph{Future Generation Computer Systems}}
  \bibinfo{volume}{29}, \bibinfo{number}{7} (\bibinfo{year}{2013}),
  \bibinfo{pages}{1645--1660}.
\newblock
\showISBNx{0167739X}
\showISSN{0167739X}
\urldef\tempurl%
\url{https://doi.org/10.1016/j.future.2013.01.010}
\showDOI{\tempurl}
\showeprint[arxiv]{1207.0203}


\bibitem[\protect\citeauthoryear{Han, Lee, Crespi, Heo, {Van Luong}, Brut, and
  Gatellier}{Han et~al\mbox{.}}{2014}]%
        {Han2014}
\bibfield{author}{\bibinfo{person}{Son~N. Han}, \bibinfo{person}{Gyu~Myoung
  Lee}, \bibinfo{person}{Noel Crespi}, \bibinfo{person}{Kyongwoo Heo},
  \bibinfo{person}{Nguyen {Van Luong}}, \bibinfo{person}{Mihaela Brut}, {and}
  \bibinfo{person}{Patrick Gatellier}.} \bibinfo{year}{2014}\natexlab{}.
\newblock \showarticletitle{{DPWSim: A simulation toolkit for IoT applications
  using devices profile for web services}}.
\newblock \bibinfo{journal}{\emph{2014 IEEE World Forum on Internet of Things,
  WF-IoT 2014}} (\bibinfo{year}{2014}), \bibinfo{pages}{544--547}.
\newblock
\showISBNx{9781479934591}
\urldef\tempurl%
\url{https://doi.org/10.1109/WF-IoT.2014.6803226}
\showDOI{\tempurl}


\bibitem[\protect\citeauthoryear{Hanes, Salgueiro, Grossetete, Barton, and
  Henry}{Hanes et~al\mbox{.}}{[n. d.]}]%
        {book:1689191}
\bibfield{author}{\bibinfo{person}{David Hanes}, \bibinfo{person}{Gonzalo
  Salgueiro}, \bibinfo{person}{Patrick Grossetete}, \bibinfo{person}{Rob
  Barton}, {and} \bibinfo{person}{Jerome Henry}.} \bibinfo{year}{[n.
  d.]}\natexlab{}.
\newblock \bibinfo{booktitle}{\emph{IoT Fundamentals: Networking Technologies,
  Protocols, and Use Cases for the Internet of Things}}.
\newblock


\bibitem[\protect\citeauthoryear{Hanmer}{Hanmer}{2013}]%
        {hanmer2013patterns}
\bibfield{author}{\bibinfo{person}{Robert Hanmer}.}
  \bibinfo{year}{2013}\natexlab{}.
\newblock \bibinfo{booktitle}{\emph{Patterns for fault tolerant software}}.
\newblock \bibinfo{publisher}{John Wiley \& Sons}.
\newblock


\bibitem[\protect\citeauthoryear{Hao, Morin, and Berre}{Hao
  et~al\mbox{.}}{2012}]%
        {Hao2012}
\bibfield{author}{\bibinfo{person}{Runze Hao}, \bibinfo{person}{Brice Morin},
  {and} \bibinfo{person}{Arne-J{\o}rgen Berre}.}
  \bibinfo{year}{2012}\natexlab{}.
\newblock \showarticletitle{{A semi-automatic behavioral mediation approach
  based on models@runtime}}.
\newblock \bibinfo{journal}{\emph{Mrt@Runtime}} (\bibinfo{year}{2012}),
  \bibinfo{pages}{67--71}.
\newblock
\showISBNx{9781450318020}
\urldef\tempurl%
\url{https://doi.org/10.1145/2422518.2422529}
\showDOI{\tempurl}


\bibitem[\protect\citeauthoryear{Harrand, Fleurey, Morin, and Husa}{Harrand
  et~al\mbox{.}}{2016}]%
        {Harrand:2016:TLC:2976767.2976812}
\bibfield{author}{\bibinfo{person}{Nicolas Harrand}, \bibinfo{person}{Franck
  Fleurey}, \bibinfo{person}{Brice Morin}, {and} \bibinfo{person}{Knut~Eilif
  Husa}.} \bibinfo{year}{2016}\natexlab{}.
\newblock \showarticletitle{ThingML: A Language and Code Generation Framework
  for Heterogeneous Targets}.
\newblock  (\bibinfo{year}{2016}), \bibinfo{pages}{125--135}.
\newblock
\showISBNx{978-1-4503-4321-3}
\urldef\tempurl%
\url{https://doi.org/10.1145/2976767.2976812}
\showDOI{\tempurl}


\bibitem[\protect\citeauthoryear{Herzberg, Bekerman, and Zeifman}{Herzberg
  et~al\mbox{.}}{2016}]%
        {Herzberg2016Oct}
\bibfield{author}{\bibinfo{person}{Ben Herzberg}, \bibinfo{person}{Dima
  Bekerman}, {and} \bibinfo{person}{Igal Zeifman}.}
  \bibinfo{year}{2016}\natexlab{}.
\newblock \showarticletitle{{Breaking Down Mirai: An IoT DDoS Botnet
  Analysis}}.
\newblock \bibinfo{journal}{\emph{Incapsula Blog}} (\bibinfo{date}{Oct}
  \bibinfo{year}{2016}).
\newblock
\urldef\tempurl%
\url{https://www.incapsula.com/blog/malware-analysis-mirai-ddos-botnet.html}
\showURL{%
\tempurl}


\bibitem[\protect\citeauthoryear{Hossain, Fotouhi, and Hasan}{Hossain
  et~al\mbox{.}}{2015}]%
        {Hossain2015}
\bibfield{author}{\bibinfo{person}{Md.~Mahmud Hossain}, \bibinfo{person}{Maziar
  Fotouhi}, {and} \bibinfo{person}{Ragib Hasan}.}
  \bibinfo{year}{2015}\natexlab{}.
\newblock \showarticletitle{{Towards an Analysis of Security Issues,
  Challenges, and Open Problems in the Internet of Things}}.
\newblock \bibinfo{journal}{\emph{2015 IEEE World Congress on Services}}
  (\bibinfo{year}{2015}), \bibinfo{pages}{21--28}.
\newblock
\showISBNx{978-1-4673-7275-6}
\showISSN{2378-3818}
\urldef\tempurl%
\url{https://doi.org/10.1109/SERVICES.2015.12}
\showDOI{\tempurl}


\bibitem[\protect\citeauthoryear{Hurlburt, Voas, and Miller}{Hurlburt
  et~al\mbox{.}}{2012}]%
        {Hurlburt2012}
\bibfield{author}{\bibinfo{person}{G~F Hurlburt}, \bibinfo{person}{J Voas},
  {and} \bibinfo{person}{K~W Miller}.} \bibinfo{year}{2012}\natexlab{}.
\newblock \showarticletitle{{The Internet of Things: A Reality Check}}.
\newblock \bibinfo{journal}{\emph{IT Professional}} \bibinfo{volume}{14},
  \bibinfo{number}{June} (\bibinfo{year}{2012}), \bibinfo{pages}{56--59}.
\newblock
\showISBNx{1520-9202 VO - 14}
\showISSN{15209202}
\urldef\tempurl%
\url{https://doi.org/10.1109/MITP.2012.60}
\showDOI{\tempurl}


\bibitem[\protect\citeauthoryear{IEEE}{IEEE}{1990}]%
        {159342}
\bibfield{author}{\bibinfo{person}{IEEE}.} \bibinfo{year}{1990}\natexlab{}.
\newblock \showarticletitle{{IEEE Standard Glossary of Software Engineering
  Terminology}}.
\newblock \bibinfo{journal}{\emph{IEEE Std 610.12-1990}} (\bibinfo{date}{Dec}
  \bibinfo{year}{1990}), \bibinfo{pages}{1--84}.
\newblock
\urldef\tempurl%
\url{https://doi.org/10.1109/IEEESTD.1990.101064}
\showDOI{\tempurl}


\bibitem[\protect\citeauthoryear{Iorga, Feldman, Barton, Martin, Goren, and
  Mahmoudi}{Iorga et~al\mbox{.}}{2018}]%
        {iorga2018fog}
\bibfield{author}{\bibinfo{person}{Michaela Iorga}, \bibinfo{person}{Larry
  Feldman}, \bibinfo{person}{Robert Barton}, \bibinfo{person}{Michael~J
  Martin}, \bibinfo{person}{Nedim~S Goren}, {and} \bibinfo{person}{Charif
  Mahmoudi}.} \bibinfo{year}{2018}\natexlab{}.
\newblock \bibinfo{booktitle}{\emph{Fog Computing Conceptual Model}}.
\newblock \bibinfo{type}{{T}echnical {R}eport}.
\newblock


\bibitem[\protect\citeauthoryear{{ISO/IEC JTC 1}}{{ISO/IEC JTC 1}}{2014}]%
        {isodef}
\bibfield{author}{\bibinfo{person}{{ISO/IEC JTC 1}}.}
  \bibinfo{year}{2014}\natexlab{}.
\newblock \bibinfo{booktitle}{\emph{Internet of Things (IoT) - Preliminary
  Report}}.
\newblock \bibinfo{type}{{T}echnical {R}eport}. \bibinfo{institution}{ISO}.
\newblock


\bibitem[\protect\citeauthoryear{Janes}{Janes}{2017}]%
        {iotdbs}
\bibfield{author}{\bibinfo{person}{D. Janes}.} \bibinfo{year}{2017}\natexlab{}.
\newblock \bibinfo{booktitle}{\emph{IOTDB}}.
\newblock \bibinfo{type}{{T}echnical {R}eport}.
  \bibinfo{institution}{IOTDB.org}.
\newblock


\bibitem[\protect\citeauthoryear{Jennings, Shelby, Cisco, Sensinode, Ericsson,
  and Arkko}{Jennings et~al\mbox{.}}{2013}]%
        {SENML2018}
\bibfield{author}{\bibinfo{person}{C. Jennings}, \bibinfo{person}{Z. Shelby},
  \bibinfo{person}{Cisco}, \bibinfo{person}{Sensinode},
  \bibinfo{person}{Ericsson}, {and} \bibinfo{person}{J. Arkko}.}
  \bibinfo{year}{2013}\natexlab{}.
\newblock \bibinfo{title}{Media Types for Sensor Markup Language (SENML)}.
\newblock
\newblock
\urldef\tempurl%
\url{https://tools.ietf.org/html/draft-jennings-senml-10}
\showURL{%
\tempurl}
\newblock
\shownote{[Online; accessed 19. May 2018].}


\bibitem[\protect\citeauthoryear{Jiang and Hassan}{Jiang and Hassan}{2015}]%
        {7123673}
\bibfield{author}{\bibinfo{person}{Z.~M. Jiang} {and} \bibinfo{person}{A.~E.
  Hassan}.} \bibinfo{year}{2015}\natexlab{}.
\newblock \showarticletitle{A Survey on Load Testing of Large-Scale Software
  Systems}.
\newblock \bibinfo{journal}{\emph{IEEE Transactions on Software Engineering}}
  \bibinfo{volume}{41}, \bibinfo{number}{11} (\bibinfo{date}{Nov}
  \bibinfo{year}{2015}), \bibinfo{pages}{1091--1118}.
\newblock
\showISSN{0098-5589}
\urldef\tempurl%
\url{https://doi.org/10.1109/TSE.2015.2445340}
\showDOI{\tempurl}


\bibitem[\protect\citeauthoryear{Johnston, Hanna, and Millar}{Johnston
  et~al\mbox{.}}{2004}]%
        {Johnston2004}
\bibfield{author}{\bibinfo{person}{Wesley~M. Johnston},
  \bibinfo{person}{J.~R.~Paul Hanna}, {and} \bibinfo{person}{Richard~J.
  Millar}.} \bibinfo{year}{2004}\natexlab{}.
\newblock \showarticletitle{{Advances in dataflow programming languages}}.
\newblock \bibinfo{journal}{\emph{Comput. Surveys}} \bibinfo{volume}{36},
  \bibinfo{number}{1} (\bibinfo{year}{2004}), \bibinfo{pages}{1--34}.
\newblock
\showISBNx{0360-0300}
\showISSN{03600300}
\urldef\tempurl%
\url{https://doi.org/10.1145/1013208.1013209}
\showDOI{\tempurl}


\bibitem[\protect\citeauthoryear{Kafle, Fukushima, and Harai}{Kafle
  et~al\mbox{.}}{2016}]%
        {Kafle2016}
\bibfield{author}{\bibinfo{person}{Ved~P. Kafle}, \bibinfo{person}{Yusuke
  Fukushima}, {and} \bibinfo{person}{Hiroaki Harai}.}
  \bibinfo{year}{2016}\natexlab{}.
\newblock \showarticletitle{{Internet of things standardization in ITU and
  prospective networking technologies}}.
\newblock \bibinfo{journal}{\emph{IEEE Communications Magazine}}
  \bibinfo{volume}{54}, \bibinfo{number}{9} (\bibinfo{date}{sep}
  \bibinfo{year}{2016}), \bibinfo{pages}{43--49}.
\newblock
\showISBNx{0163-6804}
\showISSN{0163-6804}
\urldef\tempurl%
\url{https://doi.org/10.1109/MCOM.2016.7565271}
\showDOI{\tempurl}


\bibitem[\protect\citeauthoryear{Kevin}{Kevin}{2009}]%
        {AshtonKevin2009}
\bibfield{author}{\bibinfo{person}{Ashton Kevin}.}
  \bibinfo{year}{2009}\natexlab{}.
\newblock \showarticletitle{{That 'Internet of Things' Thing}}.
\newblock \bibinfo{journal}{\emph{RFID journal}} (\bibinfo{year}{2009}).
\newblock


\bibitem[\protect\citeauthoryear{Kirichek and Koucheryavy}{Kirichek and
  Koucheryavy}{2016}]%
        {Kirichek2016}
\bibfield{author}{\bibinfo{person}{Ruslan Kirichek} {and}
  \bibinfo{person}{Andrey Koucheryavy}.} \bibinfo{year}{2016}\natexlab{}.
\newblock \bibinfo{booktitle}{\emph{Internet of Things Laboratory Test Bed}}.
\newblock \bibinfo{publisher}{Springer India}, \bibinfo{address}{New Delhi},
  \bibinfo{pages}{485--494}.
\newblock
\showISBNx{978-81-322-2580-5}
\urldef\tempurl%
\url{https://doi.org/10.1007/978-81-322-2580-5_44}
\showDOI{\tempurl}


\bibitem[\protect\citeauthoryear{Kirkland}{Kirkland}{2015}]%
        {James15}
\bibfield{author}{\bibinfo{person}{James Kirkland}.}
  \bibinfo{year}{2015}\natexlab{}.
\newblock \bibinfo{title}{{Internet of Things: insights from Red Hat - RHD
  Blog}}.
\newblock
\newblock
\urldef\tempurl%
\url{https://developers.redhat.com/blog/2015/03/31/internet-of-things-insights-from-red-hat}
\showURL{%
\tempurl}
\newblock
\shownote{[Online; accessed 22. May 2018].}


\bibitem[\protect\citeauthoryear{Koopman}{Koopman}{2011}]%
        {koopman2011embedded}
\bibfield{author}{\bibinfo{person}{Philip Koopman}.}
  \bibinfo{year}{2011}\natexlab{}.
\newblock \showarticletitle{Embedded Software Testing}.
\newblock  (\bibinfo{year}{2011}).
\newblock
\urldef\tempurl%
\url{http://www. ece. cmu. edu/\~ece649/lectures/08\_testing. pdf}
\showURL{%
\tempurl}


\bibitem[\protect\citeauthoryear{Korzun, Balandin, and Gurtov}{Korzun
  et~al\mbox{.}}{2013}]%
        {Korzun2013}
\bibfield{author}{\bibinfo{person}{Dmitry~G Korzun}, \bibinfo{person}{Sergey~I
  Balandin}, {and} \bibinfo{person}{Andrei~V Gurtov}.}
  \bibinfo{year}{2013}\natexlab{}.
\newblock \showarticletitle{{Deployment of Smart Spaces in Internet of Things:
  Overview of the Design Challenges}}.
\newblock \bibinfo{pages}{48--59}.
\newblock
\urldef\tempurl%
\url{https://doi.org/10.1007/978-3-642-40316-3_5}
\showDOI{\tempurl}


\bibitem[\protect\citeauthoryear{Leau, Loo, Tham, and Tan}{Leau
  et~al\mbox{.}}{2012}]%
        {leau2012software}
\bibfield{author}{\bibinfo{person}{Yu~Beng Leau}, \bibinfo{person}{Wooi~Khong
  Loo}, \bibinfo{person}{Wai~Yip Tham}, {and} \bibinfo{person}{Soo~Fun Tan}.}
  \bibinfo{year}{2012}\natexlab{}.
\newblock \showarticletitle{Software development life cycle AGILE vs
  traditional approaches}. In \bibinfo{booktitle}{\emph{International
  Conference on Information and Network Technology}},
  Vol.~\bibinfo{volume}{37}. \bibinfo{pages}{162--167}.
\newblock


\bibitem[\protect\citeauthoryear{Levis and Lee}{Levis and Lee}{2003}]%
        {levis2003tossim}
\bibfield{author}{\bibinfo{person}{Philip Levis} {and} \bibinfo{person}{Nelson
  Lee}.} \bibinfo{year}{2003}\natexlab{}.
\newblock \showarticletitle{Tossim: A simulator for tinyos networks}.
\newblock \bibinfo{journal}{\emph{UC Berkeley, September}}
  \bibinfo{volume}{24} (\bibinfo{year}{2003}).
\newblock


\bibitem[\protect\citeauthoryear{Lewis}{Lewis}{2016}]%
        {lewis_2016}
\bibfield{author}{\bibinfo{person}{Karen Lewis}.}
  \bibinfo{year}{2016}\natexlab{}.
\newblock \bibinfo{title}{Fantastic news for Open Source and IoT fans}.
\newblock
\newblock
\urldef\tempurl%
\url{https://www.ibm.com/blogs/internet-of-things/open-source-iot/}
\showURL{%
\tempurl}


\bibitem[\protect\citeauthoryear{Li, Azaria, and Myers}{Li
  et~al\mbox{.}}{2017a}]%
        {li2017sugilite}
\bibfield{author}{\bibinfo{person}{Toby Jia-Jun Li}, \bibinfo{person}{Amos
  Azaria}, {and} \bibinfo{person}{Brad~A Myers}.}
  \bibinfo{year}{2017}\natexlab{a}.
\newblock \showarticletitle{SUGILITE: creating multimodal smartphone automation
  by demonstration}. In \bibinfo{booktitle}{\emph{Proceedings of the 2017 CHI
  Conference on Human Factors in Computing Systems}}. ACM,
  \bibinfo{pages}{6038--6049}.
\newblock


\bibitem[\protect\citeauthoryear{Li, Li, Chen, and Myers}{Li
  et~al\mbox{.}}{2017b}]%
        {10.1007/978-3-319-58735-6_1}
\bibfield{author}{\bibinfo{person}{Toby Jia-Jun Li}, \bibinfo{person}{Yuanchun
  Li}, \bibinfo{person}{Fanglin Chen}, {and} \bibinfo{person}{Brad~A. Myers}.}
  \bibinfo{year}{2017}\natexlab{b}.
\newblock \showarticletitle{Programming IoT Devices by Demonstration Using
  Mobile Apps}. In \bibinfo{booktitle}{\emph{End-User Development}},
  \bibfield{editor}{\bibinfo{person}{Simone Barbosa}, \bibinfo{person}{Panos
  Markopoulos}, \bibinfo{person}{Fabio Patern{\`o}}, \bibinfo{person}{Simone
  Stumpf}, {and} \bibinfo{person}{Stefano Valtolina}} (Eds.).
  \bibinfo{publisher}{Springer International Publishing},
  \bibinfo{address}{Cham}, \bibinfo{pages}{3--17}.
\newblock
\showISBNx{978-3-319-58735-6}


\bibitem[\protect\citeauthoryear{Liang, Huang, and Khalafbeigi}{Liang
  et~al\mbox{.}}{2016}]%
        {ogciot}
\bibfield{author}{\bibinfo{person}{Steve Liang}, \bibinfo{person}{Chih-Yuan
  Huang}, {and} \bibinfo{person}{Tania Khalafbeigi}.}
  \bibinfo{year}{2016}\natexlab{}.
\newblock \bibinfo{booktitle}{\emph{OGC SensorThings API}}.
\newblock \bibinfo{type}{{T}echnical {R}eport}. \bibinfo{institution}{Open
  Geospatial Consortium}.
\newblock


\bibitem[\protect\citeauthoryear{Lima}{Lima}{2016}]%
        {Lima2016}
\bibfield{author}{\bibinfo{person}{Bruno Lima}.}
  \bibinfo{year}{2016}\natexlab{}.
\newblock \showarticletitle{{Automated Scenario-Based Testing of Distributed
  and Heterogeneous Systems}}.
\newblock \bibinfo{journal}{\emph{Proceedings - 2016 IEEE International
  Conference on Software Testing, Verification and Validation, ICST 2016}}
  (\bibinfo{year}{2016}), \bibinfo{pages}{383--384}.
\newblock
\showISBNx{9781509018260}
\urldef\tempurl%
\url{https://doi.org/10.1109/ICST.2016.49}
\showDOI{\tempurl}


\bibitem[\protect\citeauthoryear{Linzhang, Jiesong, Xiaofeng, Jun, Xuandong,
  and Guo}{Linzhang et~al\mbox{.}}{2004}]%
        {linzhang2004generating}
\bibfield{author}{\bibinfo{person}{Wang Linzhang}, \bibinfo{person}{Yuan
  Jiesong}, \bibinfo{person}{Yu Xiaofeng}, \bibinfo{person}{Hu Jun},
  \bibinfo{person}{Li Xuandong}, {and} \bibinfo{person}{Zheng Guo}.}
  \bibinfo{year}{2004}\natexlab{}.
\newblock \showarticletitle{{Generating test cases from UML activity diagram
  based on gray-box method}}. In \bibinfo{booktitle}{\emph{Software Engineering
  Conference, 2004. 11th Asia-Pacific}}. IEEE, \bibinfo{pages}{284--291}.
\newblock


\bibitem[\protect\citeauthoryear{Looga, Ou, Deng, and Yla-Jaaski}{Looga
  et~al\mbox{.}}{2012}]%
        {looga2012mammoth}
\bibfield{author}{\bibinfo{person}{Vilen Looga}, \bibinfo{person}{Zhonghong
  Ou}, \bibinfo{person}{Yang Deng}, {and} \bibinfo{person}{Antti Yla-Jaaski}.}
  \bibinfo{year}{2012}\natexlab{}.
\newblock \showarticletitle{Mammoth: A massive-scale emulation platform for
  internet of things}. In \bibinfo{booktitle}{\emph{Cloud Computing and
  Intelligent Systems (CCIS), 2012 IEEE 2nd International Conference on}},
  Vol.~\bibinfo{volume}{3}. IEEE, \bibinfo{pages}{1235--1239}.
\newblock


\bibitem[\protect\citeauthoryear{Meszaros and Doble}{Meszaros and
  Doble}{1997}]%
        {Meszaros1997}
\bibfield{author}{\bibinfo{person}{Gerard Meszaros} {and} \bibinfo{person}{Jim
  Doble}.} \bibinfo{year}{1997}\natexlab{}.
\newblock \showarticletitle{Pattern Languages of Program Design}.
\newblock  (\bibinfo{year}{1997}), \bibinfo{pages}{529--574}.
\newblock
\showISBNx{0-201-31011-2}
\urldef\tempurl%
\url{http://dl.acm.org/citation.cfm?id=273448.273487}
\showURL{%
\tempurl}


\bibitem[\protect\citeauthoryear{Mike~Karlesky}{Mike~Karlesky}{2018}]%
        {unity}
\bibfield{author}{\bibinfo{person}{Greg~Williams Mike~Karlesky,
  Mark~VanderVoord}.} \bibinfo{year}{2018}\natexlab{}.
\newblock \bibinfo{title}{Unity Test API}.
\newblock
\newblock
\urldef\tempurl%
\url{https://github.com/ThrowTheSwitch/Unity#unity-test-api}
\showURL{%
\tempurl}


\bibitem[\protect\citeauthoryear{Morin, Harrand, and Fleurey}{Morin
  et~al\mbox{.}}{2017}]%
        {Morin2017}
\bibfield{author}{\bibinfo{person}{Brice Morin}, \bibinfo{person}{Nicolas
  Harrand}, {and} \bibinfo{person}{Franck Fleurey}.}
  \bibinfo{year}{2017}\natexlab{}.
\newblock \showarticletitle{{Model-Based Software Engineering to Tame the IoT
  Jungle}}.
\newblock \bibinfo{journal}{\emph{IEEE Software}} \bibinfo{volume}{34},
  \bibinfo{number}{1} (\bibinfo{year}{2017}), \bibinfo{pages}{30--36}.
\newblock
\showISSN{07407459}
\urldef\tempurl%
\url{https://doi.org/10.1109/MS.2017.11}
\showDOI{\tempurl}


\bibitem[\protect\citeauthoryear{Ostrand}{Ostrand}{2002}]%
        {ostrand2002white}
\bibfield{author}{\bibinfo{person}{Thomas Ostrand}.}
  \bibinfo{year}{2002}\natexlab{}.
\newblock \showarticletitle{{White-Box Testing}}.
\newblock \bibinfo{journal}{\emph{Encyclopedia of Software Engineering}}
  (\bibinfo{year}{2002}).
\newblock


\bibitem[\protect\citeauthoryear{Perry, Blumenthal, and Hinden}{Perry
  et~al\mbox{.}}{1988}]%
        {Perry1988}
\bibfield{author}{\bibinfo{person}{Dennis~G Perry}, \bibinfo{person}{Steven~H
  Blumenthal}, {and} \bibinfo{person}{Robert~M Hinden}.}
  \bibinfo{year}{1988}\natexlab{}.
\newblock \showarticletitle{{The ARPANET and the DARPA Internet}}.
\newblock \bibinfo{journal}{\emph{Library Hi Tech}} \bibinfo{volume}{6},
  \bibinfo{number}{2} (\bibinfo{date}{feb} \bibinfo{year}{1988}),
  \bibinfo{pages}{51--62}.
\newblock
\showISSN{0737-8831}
\urldef\tempurl%
\url{https://doi.org/10.1108/eb047726}
\showDOI{\tempurl}


\bibitem[\protect\citeauthoryear{Pflanzner, Kertesz, Spinnewyn, and
  Latre}{Pflanzner et~al\mbox{.}}{2016}]%
        {Pflanzner2016}
\bibfield{author}{\bibinfo{person}{T. Pflanzner}, \bibinfo{person}{A. Kertesz},
  \bibinfo{person}{B. Spinnewyn}, {and} \bibinfo{person}{S. Latre}.}
  \bibinfo{year}{2016}\natexlab{}.
\newblock \showarticletitle{{MobIoTSim: Towards a mobile IoT device
  simulator}}.
\newblock \bibinfo{journal}{\emph{Proceedings - 2016 4th International
  Conference on Future Internet of Things and Cloud Workshops, W-FiCloud 2016}}
  (\bibinfo{year}{2016}), \bibinfo{pages}{21--27}.
\newblock
\showISBNx{9781509039463}
\urldef\tempurl%
\url{https://doi.org/10.1109/W-FiCloud.2016.21}
\showDOI{\tempurl}


\bibitem[\protect\citeauthoryear{Prehofer and Chiarabini}{Prehofer and
  Chiarabini}{2013}]%
        {Prehofer2013}
\bibfield{author}{\bibinfo{person}{Christian Prehofer} {and}
  \bibinfo{person}{Luca Chiarabini}.} \bibinfo{year}{2013}\natexlab{}.
\newblock \showarticletitle{{From IoT Mashups to Model-based IoT}}.
\newblock \bibinfo{journal}{\emph{W3C Workshop on the Web of Things}}
  (\bibinfo{year}{2013}).
\newblock


\bibitem[\protect\citeauthoryear{Prehofer and Chiarabini}{Prehofer and
  Chiarabini}{2015}]%
        {Prehofer2015}
\bibfield{author}{\bibinfo{person}{Christian Prehofer} {and}
  \bibinfo{person}{Luca Chiarabini}.} \bibinfo{year}{2015}\natexlab{}.
\newblock \showarticletitle{{From Internet of things mashups to model-based
  development}}.
\newblock \bibinfo{journal}{\emph{Proceedings - International Computer Software
  and Applications Conference}}  \bibinfo{volume}{3} (\bibinfo{year}{2015}),
  \bibinfo{pages}{499--504}.
\newblock
\showISBNx{9781467365635}
\showISSN{07303157}
\urldef\tempurl%
\url{https://doi.org/10.1109/COMPSAC.2015.263}
\showDOI{\tempurl}


\bibitem[\protect\citeauthoryear{Project}{Project}{2017}]%
        {owasp}
\bibfield{author}{\bibinfo{person}{Open Web Application~Security Project}.}
  \bibinfo{year}{2017}\natexlab{}.
\newblock \bibinfo{title}{Tester IoT Security Guidance}.
\newblock
\newblock
\urldef\tempurl%
\url{https://www.owasp.org/index.php/OWASP_Internet_of_Things_Project}
\showURL{%
\tempurl}


\bibitem[\protect\citeauthoryear{Ramadas, Domingues, Dias, Aguiar, and
  Ferreira}{Ramadas et~al\mbox{.}}{2017}]%
        {Ramadas2017}
\bibfield{author}{\bibinfo{person}{Antonio Ramadas}, \bibinfo{person}{Gil
  Domingues}, \bibinfo{person}{Joao~Pedro Dias}, \bibinfo{person}{Ademar
  Aguiar}, {and} \bibinfo{person}{Hugo~Sereno Ferreira}.}
  \bibinfo{year}{2017}\natexlab{}.
\newblock \showarticletitle{{Patterns for Things that Fail}}. In
  \bibinfo{booktitle}{\emph{Proceedings of the 24th Conference on Pattern
  Languages of Programs}} \emph{(\bibinfo{series}{PLoP '17})}.
  \bibinfo{publisher}{ACM - Association for Computing Machinery}.
\newblock


\bibitem[\protect\citeauthoryear{Ray}{Ray}{2017}]%
        {Ray2017}
\bibfield{author}{\bibinfo{person}{Partha~Pratim Ray}.}
  \bibinfo{year}{2017}\natexlab{}.
\newblock \showarticletitle{{A Survey on Visual Programming Languages in
  Internet of Things}}.
\newblock \bibinfo{journal}{\emph{Scientific Programming}}
  \bibinfo{volume}{2017} (\bibinfo{year}{2017}), \bibinfo{pages}{1--6}.
\newblock
\showISSN{1058-9244}
\urldef\tempurl%
\url{https://doi.org/10.1155/2017/1231430}
\showDOI{\tempurl}


\bibitem[\protect\citeauthoryear{Reinfurt, Breitenb{\"{u}}cher, Falkenthal,
  Leymann, and Riegg}{Reinfurt et~al\mbox{.}}{2016}]%
        {Reinfurt2016}
\bibfield{author}{\bibinfo{person}{Lukas Reinfurt}, \bibinfo{person}{Uwe
  Breitenb{\"{u}}cher}, \bibinfo{person}{Michael Falkenthal},
  \bibinfo{person}{Frank Leymann}, {and} \bibinfo{person}{Andreas Riegg}.}
  \bibinfo{year}{2016}\natexlab{}.
\newblock \showarticletitle{{Internet of things patterns}}.
\newblock \bibinfo{journal}{\emph{21st European Conference on Pattern Languages
  of Programs - EuroPlop '16}} (\bibinfo{year}{2016}), \bibinfo{pages}{1--21}.
\newblock
\showISBNx{9781450340748}


\bibitem[\protect\citeauthoryear{Reinfurt, Breitenb{\"{u}}cher, Falkenthal,
  Leymann, and Riegg}{Reinfurt et~al\mbox{.}}{2017}]%
        {Reinfurt2017}
\bibfield{author}{\bibinfo{person}{Lukas Reinfurt}, \bibinfo{person}{Uwe
  Breitenb{\"{u}}cher}, \bibinfo{person}{Michael Falkenthal},
  \bibinfo{person}{Frank Leymann}, {and} \bibinfo{person}{Andreas Riegg}.}
  \bibinfo{year}{2017}\natexlab{}.
\newblock \showarticletitle{{Internet of things patterns for devices}}. In
  \bibinfo{booktitle}{\emph{Ninth international Conferences on Pervasive
  Patterns and Applications (PATTERNS) 2017}}. \bibinfo{pages}{117--126}.
\newblock
\showISBNx{9781450340748}
\urldef\tempurl%
\url{https://doi.org/10.1145/3011784.3011789}
\showDOI{\tempurl}


\bibitem[\protect\citeauthoryear{Richardson}{Richardson}{2018}]%
        {microservicespatterns}
\bibfield{author}{\bibinfo{person}{Chris Richardson}.}
  \bibinfo{year}{2018}\natexlab{}.
\newblock \bibinfo{booktitle}{\emph{Microservices Patterns: With examples in
  Java}}.
\newblock \bibinfo{publisher}{Manning}.
\newblock


\bibitem[\protect\citeauthoryear{{Risteska Stojkoska} and
  Trivodaliev}{{Risteska Stojkoska} and Trivodaliev}{2017}]%
        {RisteskaStojkoska2017}
\bibfield{author}{\bibinfo{person}{Biljana~L. {Risteska Stojkoska}} {and}
  \bibinfo{person}{Kire~V. Trivodaliev}.} \bibinfo{year}{2017}\natexlab{}.
\newblock \showarticletitle{{A review of Internet of Things for smart home:
  Challenges and solutions}}.
\newblock \bibinfo{journal}{\emph{Journal of Cleaner Production}}
  \bibinfo{volume}{140} (\bibinfo{year}{2017}), \bibinfo{pages}{1454--1464}.
\newblock
\showISSN{09596526}
\urldef\tempurl%
\url{https://doi.org/10.1016/j.jclepro.2016.10.006}
\showDOI{\tempurl}


\bibitem[\protect\citeauthoryear{Riungu, Taipale, and Smolander}{Riungu
  et~al\mbox{.}}{2010}]%
        {riungu2010research}
\bibfield{author}{\bibinfo{person}{Leah~Muthoni Riungu}, \bibinfo{person}{Ossi
  Taipale}, {and} \bibinfo{person}{Kari Smolander}.}
  \bibinfo{year}{2010}\natexlab{}.
\newblock \showarticletitle{Research issues for software testing in the cloud}.
  In \bibinfo{booktitle}{\emph{Cloud Computing Technology and Science
  (CloudCom), 2010 IEEE Second International Conference on}}. IEEE,
  \bibinfo{pages}{557--564}.
\newblock


\bibitem[\protect\citeauthoryear{{S. K. Chang}}{{S. K. Chang}}{2002}]%
        {S.K.Chang202}
\bibfield{author}{\bibinfo{person}{{S. K. Chang}}.}
  \bibinfo{year}{2002}\natexlab{}.
\newblock \bibinfo{booktitle}{\emph{{Handbook of Software Engineering and
  Knowledge Engineering}}}.
\newblock \bibinfo{publisher}{World Scientific Publishing Co.}
\newblock
\showISBNx{9812562737}


\bibitem[\protect\citeauthoryear{Scargill}{Scargill}{2015}]%
        {scargill_2015}
\bibfield{author}{\bibinfo{person}{Peter Scargill}.}
  \bibinfo{year}{2015}\natexlab{}.
\newblock \bibinfo{title}{Node-Red Madness}.
\newblock
\newblock
\urldef\tempurl%
\url{https://tech.scargill.net/node-red-madness/}
\showURL{%
\tempurl}


\bibitem[\protect\citeauthoryear{Scully}{Scully}{2018}]%
        {IoTAnali}
\bibfield{author}{\bibinfo{person}{Padraig Scully}.}
  \bibinfo{year}{2018}\natexlab{}.
\newblock \bibinfo{title}{{The Top 10 IoT Segments in 2018 – based on 1,600
  real IoT projects}}.
\newblock
\newblock
\urldef\tempurl%
\url{https://iot-analytics.com/top-10-iot-segments-2018-real-iot-projects/}
\showURL{%
\tempurl}
\newblock
\shownote{[Online; accessed 15. May 2018].}


\bibitem[\protect\citeauthoryear{Sicari, Rizzardi, Grieco, and
  Coen-Porisini}{Sicari et~al\mbox{.}}{2015}]%
        {Sicari2015}
\bibfield{author}{\bibinfo{person}{S. Sicari}, \bibinfo{person}{A. Rizzardi},
  \bibinfo{person}{L.~A. Grieco}, {and} \bibinfo{person}{A. Coen-Porisini}.}
  \bibinfo{year}{2015}\natexlab{}.
\newblock \showarticletitle{{Security, privacy and trust in Internet of things:
  The road ahead}}.
\newblock \bibinfo{journal}{\emph{Computer Networks}}  \bibinfo{volume}{76}
  (\bibinfo{year}{2015}), \bibinfo{pages}{146--164}.
\newblock
\showISBNx{1389-1286}
\showISSN{13891286}
\urldef\tempurl%
\url{https://doi.org/10.1016/j.comnet.2014.11.008}
\showDOI{\tempurl}
\showeprint[arxiv]{1404.7799}


\bibitem[\protect\citeauthoryear{Sotiriadis, Bessis, Asimakopoulou, and
  Mustafee}{Sotiriadis et~al\mbox{.}}{2014}]%
        {Sotiriadis2014}
\bibfield{author}{\bibinfo{person}{Stelios Sotiriadis}, \bibinfo{person}{Nik
  Bessis}, \bibinfo{person}{Eleana Asimakopoulou}, {and}
  \bibinfo{person}{Navonil Mustafee}.} \bibinfo{year}{2014}\natexlab{}.
\newblock \showarticletitle{{Towards simulating the internet of things}}.
\newblock \bibinfo{journal}{\emph{IEEE 28th International Conference on
  Advanced Information Networking and Applications Workshops}}
  (\bibinfo{year}{2014}), \bibinfo{pages}{444--448}.
\newblock
\showISBNx{9781479926527}
\urldef\tempurl%
\url{https://doi.org/10.1109/WAINA.2014.74}
\showDOI{\tempurl}


\bibitem[\protect\citeauthoryear{Sousa, Correia, and Ferreira}{Sousa
  et~al\mbox{.}}{2015}]%
        {sousa2015patterns}
\bibfield{author}{\bibinfo{person}{Tiago~Boldt Sousa},
  \bibinfo{person}{Filipe~Figueiredo Correia}, {and}
  \bibinfo{person}{Hugo~Sereno Ferreira}.} \bibinfo{year}{2015}\natexlab{}.
\newblock \showarticletitle{Patterns for software orchestration on the cloud}.
  In \bibinfo{booktitle}{\emph{Proceedings of the 22nd Conference on Pattern
  Languages of Programs}}. The Hillside Group, \bibinfo{pages}{17}.
\newblock


\bibitem[\protect\citeauthoryear{Stankovic}{Stankovic}{2014}]%
        {Stankovic}
\bibfield{author}{\bibinfo{person}{John~A Stankovic}.}
  \bibinfo{year}{2014}\natexlab{}.
\newblock \showarticletitle{{Research directions for the internet of things}}.
\newblock \bibinfo{journal}{\emph{IEEE Internet of Things Journal}}
  \bibinfo{volume}{1}, \bibinfo{number}{1} (\bibinfo{year}{2014}),
  \bibinfo{pages}{3--9}.
\newblock
\showISBNx{2327-4662 VO - 1}
\showISSN{23274662 (ISSN)}
\urldef\tempurl%
\url{https://doi.org/10.1109/JIOT.2014.2312291}
\showDOI{\tempurl}


\bibitem[\protect\citeauthoryear{Sundmaeker, Guillemin, Friess, and
  Woelffle}{Sundmaeker et~al\mbox{.}}{2010}]%
        {eu10}
\bibfield{author}{\bibinfo{person}{H. Sundmaeker}, \bibinfo{person}{P.
  Guillemin}, \bibinfo{person}{P. Friess}, {and} \bibinfo{person}{S.
  Woelffle}.} \bibinfo{year}{2010}\natexlab{}.
\newblock \bibinfo{booktitle}{\emph{Vision and Challenges for Realising the
  Internet of Things}}.
\newblock \bibinfo{type}{{T}echnical {R}eport}. \bibinfo{institution}{European
  Commission}.
\newblock


\bibitem[\protect\citeauthoryear{Thomas and Barry}{Thomas and Barry}{2003}]%
        {Thomas:2003:MDD:949344.949346}
\bibfield{author}{\bibinfo{person}{Dave Thomas} {and} \bibinfo{person}{Brian~M.
  Barry}.} \bibinfo{year}{2003}\natexlab{}.
\newblock \showarticletitle{Model Driven Development: The Case for Domain
  Oriented Programming}.
\newblock  (\bibinfo{year}{2003}), \bibinfo{pages}{2--7}.
\newblock
\showISBNx{1-58113-751-6}
\urldef\tempurl%
\url{https://doi.org/10.1145/949344.949346}
\showDOI{\tempurl}


\bibitem[\protect\citeauthoryear{Trifa, Guinard, and Carrera}{Trifa
  et~al\mbox{.}}{2017}]%
        {webthingw3c}
\bibfield{author}{\bibinfo{person}{V. Trifa}, \bibinfo{person}{D. Guinard},
  {and} \bibinfo{person}{D. Carrera}.} \bibinfo{year}{2017}\natexlab{}.
\newblock \bibinfo{booktitle}{\emph{Web Thing Model}}.
\newblock \bibinfo{type}{{T}echnical {R}eport}.
  \bibinfo{institution}{EVRYTHNG}.
\newblock


\bibitem[\protect\citeauthoryear{Whitmore, Agarwal, and Da~Xu}{Whitmore
  et~al\mbox{.}}{2015}]%
        {whitmore2015internet}
\bibfield{author}{\bibinfo{person}{Andrew Whitmore}, \bibinfo{person}{Anurag
  Agarwal}, {and} \bibinfo{person}{Li Da~Xu}.} \bibinfo{year}{2015}\natexlab{}.
\newblock \showarticletitle{The Internet of Things—A survey of topics and
  trends}.
\newblock \bibinfo{journal}{\emph{Information Systems Frontiers}}
  \bibinfo{volume}{17}, \bibinfo{number}{2} (\bibinfo{year}{2015}),
  \bibinfo{pages}{261--274}.
\newblock


\bibitem[\protect\citeauthoryear{Younis and Frey}{Younis and Frey}{2003}]%
        {Younis2003}
\bibfield{author}{\bibinfo{person}{M~Bani Younis} {and} \bibinfo{person}{G
  Frey}.} \bibinfo{year}{2003}\natexlab{}.
\newblock \showarticletitle{{Formalization of Existing PLC Programs: A
  Survey}}.
\newblock \bibinfo{journal}{\emph{CESA 2003, Lille (France), Paper No. S2-R}}
  (\bibinfo{year}{2003}), \bibinfo{pages}{--00--0239}.
\newblock


\bibitem[\protect\citeauthoryear{Zeng, Garg, Strazdins, Jayaraman,
  Georgakopoulos, and Ranjan}{Zeng et~al\mbox{.}}{2017}]%
        {Zeng2017}
\bibfield{author}{\bibinfo{person}{Xuezhi Zeng}, \bibinfo{person}{Saurabh~Kumar
  Garg}, \bibinfo{person}{Peter Strazdins}, \bibinfo{person}{Prem~Prakash
  Jayaraman}, \bibinfo{person}{Dimitrios Georgakopoulos}, {and}
  \bibinfo{person}{Rajiv Ranjan}.} \bibinfo{year}{2017}\natexlab{}.
\newblock \showarticletitle{{IOTSim: A simulator for analysing IoT
  applications}}.
\newblock \bibinfo{journal}{\emph{Journal of Systems Architecture}}
  \bibinfo{volume}{72} (\bibinfo{year}{2017}), \bibinfo{pages}{93--107}.
\newblock
\showISBNx{13837621}
\showISSN{13837621}
\urldef\tempurl%
\url{https://doi.org/10.1016/j.sysarc.2016.06.008}
\showDOI{\tempurl}


\bibitem[\protect\citeauthoryear{Zhang, Cho, Wang, Hsu, Chen, and Shieh}{Zhang
  et~al\mbox{.}}{2014}]%
        {6978614}
\bibfield{author}{\bibinfo{person}{Z.~K. Zhang}, \bibinfo{person}{M.~C.~Y.
  Cho}, \bibinfo{person}{C.~W. Wang}, \bibinfo{person}{C.~W. Hsu},
  \bibinfo{person}{C.~K. Chen}, {and} \bibinfo{person}{S. Shieh}.}
  \bibinfo{year}{2014}\natexlab{}.
\newblock \showarticletitle{IoT Security: Ongoing Challenges and Research
  Opportunities}. In \bibinfo{booktitle}{\emph{2014 IEEE 7th International
  Conference on Service-Oriented Computing and Applications}}.
  \bibinfo{pages}{230--234}.
\newblock
\showISSN{2163-2871}
\urldef\tempurl%
\url{https://doi.org/10.1109/SOCA.2014.58}
\showDOI{\tempurl}


\end{thebibliography}

\end{document}